\documentclass{elsarticle} 

\usepackage{amssymb}
\usepackage{amsthm}
\usepackage{mathtools}
\usepackage{graphicx}
\usepackage{hyperref}
\usepackage{tikz}
\usepackage{xcolor}
\usepackage{stmaryrd}
\usepackage[size=scriptsize, color=yellow]{todonotes}
\usepackage{tikz}
\usetikzlibrary{arrows,decorations.pathmorphing,snakes,backgrounds,positioning,fit}
\usetikzlibrary{automata}
\hypersetup{
	colorlinks,
	linkcolor={blue!70!black},
	citecolor={blue!70!black},
	urlcolor={blue!70!black}
}
\usepackage{thm-restate}

\DeclareMathOperator{\com}{Com}
\DeclareMathOperator{\rng}{rng}
\DeclareMathOperator{\dom}{dom}

\newcommand{\pair}[2]{\ensuremath{\langle\!\langle {#1}, {#2} \rangle\!\rangle}}
\newcommand{\eqd}{\triangleq}
\newcommand{\ifd}{\stackrel{{\mbox{\tiny\ensuremath{\triangle}}}}{\Longleftrightarrow}}\newcommand{\tuple}[1]{\langle{#1}\rangle}
\newcommand{\Id}{\textsc{Id}}
\def\grasse#1{\llbracket #1 \rrbracket}

\theoremstyle{plain}
\newtheorem{theorem}{Theorem}[section]
\newtheorem{corollary}[theorem]{Corollary}
\newtheorem{lemma}[theorem]{Lemma}
\newtheorem{proposition}[theorem]{Proposition}
\theoremstyle{definition}
\newtheorem{definition}[theorem]{Definition}
\newtheorem{example}[theorem]{Example}
\newtheorem{assumption}[theorem]{Assumption}

\journal{Information and Computation}

\begin{document}
	
	\begin{frontmatter}
		
		\title{Intensional Kleene and Rice Theorems\\ for Abstract Program Semantics}
		
		\author[1]{Paolo Baldan}
		\author[1]{Francesco Ranzato}
		
		\affiliation[1]{organization={Dipartimento di Matematica,  University of Padova},%Department and Organization
			%addressline={}, 
			%city={},
			%postcode={}, 
			%state={},
			country={Italy}}

		%\affiliation[2]{organization={Dipartimento di Matematica,  University of Padova},%Department and Organization
			%addressline={}, 
			%city={},
			%postcode={}, 
			%state={},
		%	country={Italy}}
		
		\author[2]{Linpeng Zhang}
		
		\affiliation[2]{organization={Department of Computer Science, University College London},%Department and Organization
			%addressline={}, 
			%city={},
			%postcode={}, 
			%state={},
			country={UK}}
		
		\begin{abstract}
			Classical results in computability theory, notably Rice's theorem,
			focus on the extensional content of programs, namely, on the partial
			recursive functions that programs compute. 
			Later and more recent work investigated intensional
			generalisations of such results that take into
			account the way in which functions are computed, thus affected 
			by the specific programs computing them. 
			In this paper, we single out a novel class of program semantics based on 
			abstract domains of program properties
			that are able to capture nonextensional aspects of program computations,
			such as their asymptotic complexity or logical invariants, and allow us to generalise
			some foundational computability results such as Rice's Theorem and 
			Kleene's Second Recursion Theorem to these semantics. 
			In particular, it turns out that for this class of abstract program semantics, any
			nontrivial abstract property is undecidable and every decidable
			over-approximation necessarily includes an infinite set of false positives
			which covers all the values of the semantic abstract domain.
		\end{abstract}
		
		%%Graphical abstract
		%\begin{graphicalabstract}
		%\includegraphics{grabs}
		%\end{graphicalabstract}
		
		%%Research highlights
		%\begin{highlights}
		%	\item Research highlight 1
		%	\item Research highlight 2
		%\end{highlights}
		
		\begin{keyword}
			%% keywords here, in the form: keyword \sep keyword
			Computability Theory\sep Recursive Function\sep Rice's Theorem \sep Kleene's Second Recursion Theorem\sep Program Analysis\sep Affine Program Invariants
			%% PACS codes here, in the form: \PACS code \sep code
			
			%% MSC codes here, in the form: \MSC code \sep code
			%% or \MSC[2008] code \sep code (2000 is the default)
			
		\end{keyword}
		
	\end{frontmatter}
	
	%% \linenumbers
	
	\section{Introduction}
	
	Most classical results in computability theory focus on the so-called
	\emph{extensional} properties of programs, i.e., on the properties of the
	partial functions they compute. Notably, the renowned Rice's
	Theorem~\cite{Rice53} states that any nontrivial extensional
	property of programs is
	undecidable.
	Despite being very general, Rice's Theorem and similar results in
	computability theory, due to the requirement of extensionality, leave
	out several \emph{intensional} properties which are of utmost importance in the
	practice of programming. Essential intensional properties of programs include 
	their asymptotic complexity of computation, their logical invariants (e.g., 
	relations between 
	variables at program points), or any event that might happen during the execution of the program while not affecting its output.
	
	\paragraph*{State-of-the-Art}
	A generalisation of well-established results of computability theory to the realm of
	program complexity has been put forward by Asperti~\cite{Asp08}. A first
	observation is that Blum's complexity classes~\cite{Blum67}, i.e.,
	sets of recursive functions (rather than sets of programs) with some given (lower or upper) 
	bound on their (space and/or time) 
	complexity, are not adequate for investigating the decidability aspects
	of program complexity: in fact, viewed as program properties they are
	trivially extensional. Thus, a key idea in \cite{Asp08} is to focus on the so-called
	\emph{complexity cliques}, namely, sets of programs (i.e., program indices) closed with respect to their extensional input/output behaviour and their asymptotic
	complexity. Asperti~\cite{Asp08} showed how this approach enables 
	intensional versions of Rice's theorem, Rice-Shapiro theorem, and
	Kleene's second recursion theorem (\cite{Cut80,Rog67} are standard references for these foundational results) for complexity
	cliques. 
	
	More recently, a different approach has been considered by Moyen and
	Simonsen in~\cite{MS19}, where the classical definition
	of extensionality has been weakened to a notion of \emph{partial extensionality}. Roughly, a given set of programs is partially extensional if it
	includes the set of all programs computing a given partial recursive function. It is
	shown in \cite{MS19} that if a set of programs and its complement are partially extensional, then
	they cannot be recursive. Interestingly, this result can be further
	generalised by replacing the extensionality with an equivalence relation
	on programs satisfying some suitable structural conditions, 
	notably, the existence of a
	so-called intricated switching family. Moyen and 
	Simonsen~\cite{MS19} show how to derive within their framework intensional
	versions of Rice's Theorem --- generalising Asperti's result~\cite{Asp08} ---
	and Rice-Shapiro Theorem.

	\paragraph*{Main Contributions}
	Along the lines traced by Asperti~\cite{Asp08}, we investigate 
	whether and how some fundamental extensional results of
	computability theory can be systematically generalised 
	to intensional aspects of computation, but rather than focussing 
	on specific intensional properties we deal with generic \emph{abstract program semantics}. 
	More in detail, we distill two fundamental properties of abstract
	program semantics in our approach: the \emph{strong smn property}
	and the existence of a \emph{universal fair program},
	roughly, an interpreter that preserves the abstract semantics. We show
	that for abstract semantics satisfying the strong smn property and admitting a universal
	fair program, a generalisation of Kleene's second recursion theorem can
	be proved. This, in turn, leads to a generalisation of Rice's
	theorem. Besides relying on a general abstract program semantics, inspired
	by Moyen and Simonsen's approach~\cite{MS19}, we also relax the extensionality condition to partial
	extensionality. This weakening provides stronger impossibility
	results as it allows us to show that every decidable
	over-approximation necessarily contains an infinite set of false positives
	which covers all the values of the underlying semantic abstract domain.
	%to show that it is undecidable whether a given
	%program \emph{can} have a particular semantics, i.e., even nontrivial
	%overapproximations of such properties are undecidable.\todo{P: Mi sembra poco chiaro. Rifraserei come nell'abstract ``... it allows to show thavt every decidable
	%overapproximation necessarily includes an infinite set of false positives
	%which covers all values of the semantic abstract domain.}
	%
	On a different route, we establish a precise connection with Moyen and
	Simonsen's work~\cite{MS19} by showing that for
	any abstract program  semantics satisfying the strong smn property and
	a structural \emph{branching} condition (roughly, expressing some form of conditional choice), 
	we can prove the existence of an intricated switching family, which turns out to be 
	the crucial hypothesis in \cite{MS19} for deriving an 
	intensional version of Rice's theorem.

	Therefore, on the one hand, we generalise the results in~\cite{Asp08},
	going beyond complexity cliques, and, on the other hand, we provide an
	explicit characterisation of a class of program semantics that admit
	intricated switching families so that the results in~\cite{MS19} can
	be applied.
	
	Finally, we show some applications of our intensional Rice's theorem that generalise some undecidability results for intensional properties used in static program analysis. In particular, we focus on program analysis in Karr's abstract domain of
	affine relations between program variables~\cite{karr76}. By exploiting an acute reduction
	to the undecidable Post correspondence problem,
	M\"{u}ller-Olm and Seidl~\cite{MOS04Affine} prove that
	for affine programs with positive affine
	guards it is undecidable 
	whether a given nontrivial affine relation holds at a given program point or not. Here, we first show that
	this class of affine programs with positive affine
	guards, modeled as control flow graphs, 
	turns out to be Turing complete since, by  selecting
	a suitable program semantics, these programs can simulate a URM. Then, this allows us to derive the undecidability result
	in \cite{MOS04Affine} as a consequence of our results. 
	
	\paragraph*{Outline}
	The rest of the paper is structured as follows. In
	Section~\ref{se:basics}, we provide some background and our basic
	notions. 
	In Section~\ref{se:smn-fair}, we introduce the strong smn property, 
	fair universal programs, and the branching condition that will play a fundamental role in our results. 
	In Section~\ref{se:second-recursion}, we provide our generalisation of Kleene's
	second recursion theorem and use it to derive our intensional Rice's theorem. We also establish an explicit connection with the notion of intricated switching family given in~\cite{MS19}. 
	Section~\ref{se:cfg} provides some applications of our results to the 
	analysis of affine programs. 
	Section~\ref{se:related} discusses in detail the relation with some of Asperti's results \cite{Asp08} and with Rogers' systems of indices~\cite{Rog58,Rog67}. 
	Finally, Section~\ref{se:conclusions} concludes and outlines some directions of future work.
	This is a full and revised version of the conference paper~\cite{icalp21}.

	\section{Basic Notions}
	\label{se:basics}
	
	Given an $n$-ary partial function $f : \mathbb{N}^n \rightarrow \mathbb{N}$, we
	denote by $\dom(f)$ the domain of $f$ and by
	$\rng(f)\eqd \{f(\Vec{x}):\Vec{x}\in \dom(f)\}$ its range. We write
	$f(\Vec{x})\!\downarrow$ if $\Vec{x}\in \dom(f)$ and $f(\Vec{x})\!\uparrow$
	if $\Vec{x}\notin \dom(f)$. Moreover, $\lambda \Vec{x}.\!\uparrow$ denotes the always undefined function. 
	We denote by $\mathcal{F}_n\eqd \mathbb{N}^n\rightarrow\mathbb{N}$ the class of all
	$n$-ary (possibly partial) functions and by
	$\mathcal{F}\eqd \bigcup_n \mathcal{F}_n$ the class of all such functions.
	Additionally, $\mathcal{C}_n\subseteq  \mathcal{F}_n$ denotes
	the subset of $n$-ary partial recursive functions ($\mathcal{C}$ stands for computable)
	and $\mathcal{C} \eqd\bigcup_n \mathcal{C}_n$ 
	the set of all partial recursive 
	functions.
	
	\begin{assumption}[{\bf Turing completeness}]
		\label{ass-tc}
		Throughout the paper, we assume a fixed Turing complete model and we
		denote by $\mathcal{P}$ the corresponding set of programs. Moreover,
		we consider a fixed Gödel numbering for the programs in
		$\mathcal{P}$ and, given an index $a \in \mathbb{N}$, we write $P_a$
		for the $a$-th program in $\mathcal{P}$.  A program can take a
		varying number $n$ of inputs and we denote by
		$\phi_a^{(n)}\in \mathcal{C}_n$ the $n$-ary partial function
		computed by $P_a$. By Turing completeness of the model
		$\mathcal{C} = \{\phi_a^{(n)} \mid a,n\in \mathbb{N}\}$.  \qed
	\end{assumption}
	
	The binary relation between programs that compute the same $n$-ary function is called \emph{Rice's equivalence} and denoted by $\sim_{R}^n$, i.e., $$a \sim_{R}^n b \ifd \phi_a^{(n)} = \phi_b^{(n)}.$$

	Classical Rice's theorem~\cite{Rice53} compares the extension of
	programs, i.e., the functions they compute, and shows that unions of
	equivalence classes of programs computing the same function are
	undecidable. In Asperti's work~\cite{Asp08}, by relying on the notion
	of complexity clique, the asymptotic program complexity can be taken
	into account. Our idea here is to further generalise the approach
	in~\cite{Asp08} by considering generic program semantics rather than
	program complexity.  Additionally, an equivalence relation on program
	semantics allows us to further abstract and identify programs with
	different abstract semantics. This turns out to be worthwhile in many applications,
	e.g., the precise time/space
	program complexity is typically abstracted by considering asymptotic complexity classes.

	\begin{definition}[{\bf Abstract semantics}]
		\label{de:sem-fw}
		An \emph{abstract semantics} is a pair
		$\langle \pi, \equiv_\pi \rangle$ where:
		
		\begin{enumerate}[(1)]
			\item ${\pi:\mathbb{N}^2\rightarrow \mathcal{F}}$ associates a program
			index $a$ and arity $n$ with an $n$-ary function
			$\pi_a^{(n)} \in\mathcal{F}_n$, called the \emph{semantics} of $a$;
			\item $\mathbin{\equiv_\pi}\subseteq \mathcal{F}\times \mathcal{F}$ is an
			equivalence relation between functions.
		\end{enumerate} 
		
		Given $n\in\mathbb{N}$, the \emph{$n$-ary
			program equivalence} induced by an abstract semantics $\langle \pi, \equiv_\pi\rangle$
		is the equivalence
		$\mathbin{\sim_{\pi}^{n}} \subseteq \mathbb{N} \times \mathbb{N}$ defined as follows: for all $a,b \in \mathbb{N}$,
		$$
		a \sim_{\pi}^{n} b \quad \ifd \quad
		\pi_a^{(n)} \equiv_{\pi} \pi_b^{(n)}. \eqno\qed 
		$$
	\end{definition}
	
	The notation for the case of arity $n=1$ will be simplified by omitting the arity, 
	e.g., we will write $\phi_a$ and $\sim_{\pi}$ in place of $\phi_a^{(1)}$ and $\sim_{\pi}^1$, respectively. Abstract semantics can be viewed as a generalisation of the notion of system of indices (or numbering), as found in standard reference textbooks~\cite{Odi89,Rog67}. This is discussed in detail later in 
	Section~\ref{subse:indices}. Let us now show how the standard extensional interpretation of programs, complexity and complexity cliques can be cast into our setting.
	
	\begin{example}[{\bf Concrete semantics}]
		\label{ex:concrete}
		The concrete input/output semantics can be trivially seen as an
		abstract semantics $\langle \phi, = \rangle$ where $\phi_a^{(n)}$ is
		the $n$-ary function computed by $P_a$ and $=$ is the equality
		between functions. Observe that this concrete semantics induces an
		$n$-ary program equivalence which is Rice's equivalence $\sim_{R}^n$.
		\qed
	\end{example}
	
	\begin{example}[{\bf Domain semantics}]
		\label{ex:domain}
		For a given set of inputs $S \subseteq \mathbb{N}$, consider
		$\langle \phi, \equiv_S \rangle$ where $\phi_a^{(n)}$ is the $n$-ary
		function computed by $P_a$ and for
		$f, g : \mathbb{N}^n \to \mathbb{N}$, we define $f \equiv_S g$ $\ifd$
		$\dom(f) \cap S = \dom(g) \cap S$.
		\qed
	\end{example}

	\begin{example}[{\bf Blum complexity}]
		\label{ex:complexity}
		Let $\Phi:\mathbb{N}^2\rightarrow\mathcal{C}$ be a Blum complexity~\cite{Blum67}, 
		i.e., for all $a \in \mathbb{N}$ and
		$\vec{x} \in \mathbb{N}^n$, (1)~$\Phi_a^{(n)}(\Vec{x})\downarrow \;\Leftrightarrow\;
		\phi_a^{(n)}(\Vec{x})\downarrow$ holds, and (2)~for all $m\in \mathbb{N}$, 
		the predicate
		$\Phi_a^{(n)}(\Vec{x})=m$ is decidable. 
		Letting $\Theta(f)$ to
		denote the standard Big Theta complexity class of a function $f$,
		the pair
		$\langle \Phi, \equiv_\Phi\rangle$ defined by
		$$\Phi_a^{(n)} \equiv_\Phi \Phi_b^{(n)} \ifd
		\Phi_a^{(n)} \in \Theta(\Phi_b^{(n)})$$ 
		is an abstract semantics.
		\qed
	\end{example}
	
	\begin{example}[{\bf Complexity clique}]
		\label{ex:asperti}
		\emph{Complexity cliques} as defined by Asperti in~\cite{Asp08} can be viewed
		as an abstract semantics $\langle \pi, \equiv_\pi\rangle$, that we
		will refer to as the complexity clique semantics. For each arity $n$
		and program index $a$ let us define:
		$$
		\pi_a^{(n)}\eqd \lambda
		\Vec{y}.\pair{\phi_a^{(n)}(\Vec{y})}{\Phi_a^{(n)}(\Vec{y})}
		$$
		where $\pair{\_}{\_} :\mathbb{N}^2 \to \mathbb{N}$ is an effective
		bijective encoding for pairs and $\Phi:\mathbb{N}^2\rightarrow\mathcal{C}$ is a Blum complexity. The equivalence $\equiv_\pi$ is defined as follows:
		for all $a,b,n \in \mathbb{N}$, 
		$$
		\pi_a^{(n)} \equiv_\pi \pi_b^{(n)}
		\ifd
		\phi_a^{(n)} = \phi_b^{(n)} \land \Phi_a^{(n)} \equiv_\Phi \Phi_b^{(n)}. \eqno\qed
		$$
	\end{example}

	Classical Rice's theorem states the undecidabilty of extensional program properties. Following~\cite{MS19}, we parameterise extensional sets by means of 
	a generic equivalence relation.
	
	\begin{definition}[{\bf  $\sim$-extensional set}]
		Let $\mathbin{\sim}\subseteq\mathbb{N}\times\mathbb{N}$ be an
		equivalence relation between programs whose equivalence classes are
		denoted, for $a \in A$, by $[a]_{\sim}$.  A set of indices
		$A\subseteq \mathbb{N}$ is called:
		
		\begin{itemize}
			\item \emph{$\sim$-extensional} when for all $a,b\in\mathbb{N}$, if $a\in A$
			and $a \sim b$ then $b\in A$;
			
			\item \emph{partially $\sim$-extensional} when there exists
			$a\in\mathbb{N}$ such that $[a]_{\sim}\subseteq A$;
			
			\item \emph{universally $\sim$-extensional} when
			for all $a\in\mathbb{N}$, $[a]_{\sim}\cap A \neq \varnothing$. \qed
		\end{itemize}
	\end{definition}
	
	In words, a set $A$ is $\sim$-extensional if $A$ is a union of
	$\sim$-equivalence classes, partially \mbox{$\sim$-extensional} if $A$
	contains at least a whole $\sim$-equivalence class, and universally
	{$\sim$-exten\-sional} if $A$ contains at least an element from each
	$\sim$-equivalence class, i.e., its complement
	$\mathbb{N} \smallsetminus A$ is not partially $\sim$-extensional.  Notice that if
	$A$ is not trivial (i.e., $A\neq \varnothing$ and $A\neq \mathbb{N}$)
	and $\sim$-extensional then $A$ is partially $\sim$-extensional and
	not universally $\sim$-extensional.  Let us observe that
	$\sim_R$-extensionality is the standard notion of extensionality so
	that classical Rice's theorem~\cite{Rice53} states that if $A$ is
	$\sim_R$-extensional and not trivial then $A$ is not
	recursive.\footnote{In~\cite{MS19}, the term ``extensional'' is
		replaced by ``compatible'' when one refers to generic equivalence relations  $\sim$.}

	\section{Fair and Strong smn Semantics}
	\label{se:smn-fair}
	
	In this section, we identify some fundamental properties of
	abstract semantics that will be later used 
	in our intensional computability results. A first basic property stems from the fundamental smn
	theorem and intuitively amounts to requiring that the
	operation of fixing some parameters of a program 
	is effective and preserves its
	abstract semantics.

	\begin{definition}[{\bf Strong smn semantics}]
		\label{def:smnsemantics}
		An abstract semantics $\langle \pi, \equiv_\pi\rangle$ has the
		\emph{strong smn} (\emph{ssmn}) \emph{property}
		if, given $m,n \geq 1$, there exists a total
		computable function $s : \mathbb{N}^{m+2}\to \mathbb{N}$ such that
		for all $a,b \in \mathbb{N}$, $\vec{x} \in \mathbb{N}^m$:
		\begin{equation}\label{smn-condition}
			\lambda \vec{y}. \pi_a^{(n+1)}(\phi_b^{(m)}(\vec{x}), \Vec{y})
			\equiv_\pi
			\pi_{s(a,b,\vec{x})}^{(n)}.
		\end{equation}
		In such a case, the abstract semantics 
		$\langle \pi, \equiv_\pi\rangle$ is called \emph{strong smn}.
		\qed
	\end{definition}

	The above definition requires property \eqref{smn-condition} which is slightly
	stronger than one would expect. The natural generalisation of the standard smn property, in the style, e.g., of~\cite{Asp08}, would amount to asking that,
	given $m,n \geq 1$, there exists a total computable function
	${s : \mathbb{N}^{m+1}\to \mathbb{N}}$ such that for any program index
	$a \in \mathbb{N}$ and input $\vec{x} \in \mathbb{N}^m$, it holds $\lambda
	\vec{y}. \pi_a^{(m+n)}(\vec{x}, \Vec{y}) \equiv_\pi
	\pi^{(n)}_{s(a,\vec{x})}$.

	The concrete semantics $\langle \phi, =\rangle$ of Example~\ref{ex:concrete} clearly satisfies this ssmn
	property. In fact, the function
	$\lambda a, b, \vec{y}. \pi_a^{(n+1)}(\phi_b^{(m)}(\vec{x}), \Vec{y})$
	is computable (by composition, relying on the existence of universal
	functions), hence the existence of a total computable
	$s : \mathbb{N}^{m+2}\to \mathbb{N}$ such that
	$\lambda \vec{y}. \pi_a^{(n+1)}(\phi_b^{(m)}(\vec{x}), \Vec{y})
	\equiv_\pi \pi_{s(a,b,\vec{x})}^{(n)}$ holds, as prescribed by Definition~\ref{def:smnsemantics}, follows by the standard smn
	theorem. It is easily seen that the same applies to the domain semantics of Example~\ref{ex:domain}.
	
	The reason for the stronger requirement \eqref{smn-condition} in Definition~\ref{def:smnsemantics} is that, to deal with generic abstract semantics,
	a suitable smn definition needs to embody a condition on
	program composition (of $a$ and $b$ in Definition~\ref{def:smnsemantics}). Indeed, if we consider the semantics based on program complexity
	(i.e., Examples~\ref{ex:complexity} and \ref{ex:asperti}), it turns
	out that whenever they enjoy the smn property
	in~\cite[Definition~11]{Asp08} and, additionally, 
	they satisfy the linear time composition hypothesis in \cite[Section 4]{Asp08}
	relating the asymptotic complexities of a program
	composition to those of its components, then they are ssmn semantics
	according to Definition~\ref{def:smnsemantics}. More details on the relationship with Asperti's approach~\cite{Asp08} will be given later in Section~\ref{subse:asperti}.

	It is worth observing that for a ssmn abstract semantics $\langle \pi, \equiv_\pi \rangle$,
	there always exists a program whose denotation is equivalent to the always undefined function, namely, 
	\begin{equation}
	\begin{multlined}\label{prop:smn-undef-function}
		\text{for any arity } n\in\mathbb{N} \text{ there exists a program index } e_0\in \mathbb{N} 
		\text{ such that } \\\pi_{e_0}^{(n)}\equiv_\pi\lambda \Vec{y}.\!\uparrow.
	\end{multlined}
	\end{equation}

	In fact, if $b$ is a program index for the always undefined 
	function $\lambda \Vec{y}.\!\uparrow$ then, by \eqref{smn-condition}, we have that $\lambda \Vec{y}.\pi_{0}^{(n+1)}(\phi_{b}(0),\Vec{y}) = {\lambda \Vec{y}.\!\uparrow} \, \mathrel{\equiv_\pi} \pi_{s(0,b,0)}^{(n)}$, so that we can pick
	$e_0 \eqd s(0,b,0)$.
	
	It is also worth exhibiting an example of abstract semantics which is not ssmn.
	Let $\pi_a(\vec{x})$ be defined as the number of different variables accessed in a computation of the program $a$ on the input $\vec{x}$. 
	Then, let us observe that 
	the mere fact that $\pi_a$ is always a total function trivially makes the abstract semantics
	$\langle \pi, = \rangle$ non-ssmn.

	To generalise Kleene's second recursion theorem, besides the ssmn
	property, we need to postulate the existence of so-called \emph{fair
		universal programs}, namely, programs that 
	can simulate every other program w.r.t.\ 
	a given abstract semantics.
	This generalises the analogous 
	notion in~\cite[Definition 26]{Asp08}, where this simulation is
	specific to complexity cliques and must
	preserve both the computed function and its asymptotic complexity.
	
	\begin{definition}[{\bf Fair semantics}]
		\label{de:univ-prg}
		An index $u \in \mathbb{N}$ is a
		\emph{fair universal program} for an abstract semantics
		$\langle \pi, \equiv_\pi\rangle$ and an arity $n\in\mathbb{N}$ if
		for all $a \in \mathbb{N}$:
		\begin{equation*}
			\pi_a^{(n)} \equiv_\pi \lambda\Vec{y}.\pi_u^{(n+1)}(a, \Vec{y}).
		\end{equation*}
		An abstract semantics is  \emph{fair} if it admits a
		fair universal program for every arity. \qed
	\end{definition}
	
	Clearly, the concrete (Example~\ref{ex:concrete}) and domain
	(Example~\ref{ex:domain}) semantics are fair. In general, as noted
	in~\cite{Asp08}, the existence of a fair universal program may not
	only depend on the reference abstract semantics, but also on the
	underlying computational model. 
	For instance, when considering program complexity, as argued by
	Asperti~\cite[Section~6]{Asp08} by relying on some remarks by
	Blum~\cite{Blum71}, multi-tape Turing machines seem not to admit fair universal programs. By contrast, single tape Turing machines do have fair universal programs, despite the fact that this is commonly considered a folklore fact and cannot be properly quoted. 
	Hereafter, when referring to the complexity-based semantics of
	Examples~\ref{ex:complexity} and \ref{ex:asperti}, we will implicitly
	use that they are ssmn and fair semantics.

	\section{Kleene's Second Recursion Theorem and Rice's Theorem}
	\label{se:second-recursion}
	
	In this section, we show how some foundational results of computability theory can
	be extended to a general abstract semantics. The first approach
	relies on a generalisation of Kleene's second recursion
	theorem, which is then used to derive a corresponding Rice's
	theorem. A second approach consists in identifying conditions that
	ensure the existence of an intricated switching family in the sense
	of~\cite{MS19}, from which Rice's theorem also follows.
	
	\subsection{Kleene's Second Recursion Theorem}
	\label{se:kleene-rice}
	We show that Kleene's second recursion theorem holds for any 
	fair ssmn abstract semantics. This
	generalises the analogous result proved by Asperti~\cite[Section~5]{Asp08} for complexity
	cliques.
	
	% \begin{restatable}[Intensional Second Recursion Theorem]{theorem}{secondRecursion}
	\begin{theorem}[{\bf Intensional Second Recursion Theorem}]
		\label{th:second-recursion}
		Let $\langle \pi, \equiv_\pi \rangle$ be a fair ssmn abstract semantics.
		For any total computable function $h: \mathbb{N} \to \mathbb{N}$ and arity $n\in\mathbb{N}$, there exists an index
		$a \in \mathbb{N}$ such that
		$a \sim_\pi^n h(a)$.
	\end{theorem}
	
	\begin{proof}  
		Since $\langle \pi, \equiv_\pi \rangle$ is a fair semantics
		(Definition~\ref{de:univ-prg}), there exists $u, n \in \mathbb{N}$ such
		that $u$ is an abstract universal program for $n$-ary 
		functions. Hence, for all $x\in\mathbb{N}$:
		\[
		\pi_{h(\phi_x(x))}^{(n)} \equiv_\pi
		\lambda \Vec{y}.\pi_u^{(n+1)}(h(\phi_x(x)), \Vec{y}) \equiv_\pi
		\lambda \Vec{y}.\pi_u^{(n+1)}(h(\psi_U(x,x)), \Vec{y}),
		\]
		where $\psi_U$ is the standard unary universal function 
		for the concrete semantics $\phi$,
		i.e.,
		$\forall p \in \mathbb{N}.\ \lambda
		y.\psi_U(p,y)=\phi_p$. Note that
		$h\circ \lambda z.\psi_U(z,z)$ is computable by composition of
		computable functions. Hence, there exists $e$ such that
		$\phi_e=h\circ \lambda z.\psi_U(z,z)$. Since
		$\langle \pi, \equiv_\pi \rangle$ is a ssmn semantics
		(Definition~\ref{def:smnsemantics}), there exists a total
		computable function $s : \mathbb{N}^3 \to \mathbb{N}$ such
		that for all $x \in \mathbb{N}$:
		\[
		\lambda \Vec{y}.\pi_u^{(n+1)}(h(\psi_U(x,x)), \Vec{y})
		\equiv_\pi\lambda \Vec{y}.\pi_u^{(n+1)}(\phi_e(x), \Vec{y}) \equiv_\pi
		\pi_{s(u,e,x)}^{(n)}.
		\]
		Since $s$ is computable,
		by standard smn theorem,
		there exists $m\in \mathbb{N}$ such that $\phi_m = \lambda x.\ s(u,e,x)$. Hence, for all $x\in \mathbb{N}$:
		\[
		\pi_{\phi_m(x)}^{(n)} \equiv _\pi     \pi_{h(\phi_x(x))}^{(n)}.
		\]
		If we set $x=m$ we obtain:
		\[
		\pi_{\phi_m(m)}^{(n)}\equiv_\pi      \pi_{h(\phi_m(m))}^{(n)}.
		\]
		Because $\phi_m=\lambda x.\ s(u,e,x)$ is total, we can consider $a=\phi_m(m)$ and obtain:
		\[
		\pi_{a}^{(n)}  \equiv_\pi       \pi_{h(a)}^{(n)}
		\]
		which amounts to $a \sim_{\pi}^n h(a)$.
	\end{proof}
	
	As an example, this result, instantiated to the complexity semantics of Example~\ref{ex:complexity}, entails the impossibility of 
	designing a program transform that systematically modifies the asymptotic complexity of every program, even without preserving its input-output behavior. The details are discussed below.
	
	\begin{example}[{\bf Fixpoints of Blum complexity semantics}]
		\label{ex:complexity-fixpoint}
		Let $\tuple{\Phi,\equiv_\Phi}$ be the Blum complexity semantics of
		Example~\ref{ex:complexity}. A program transform
		$h: \mathbb{N} \to \mathbb{N}$ is a total computable function which
		maps indices of programs into indices of transformed programs. By
		applying Theorem~\ref{th:second-recursion}, for any arity
		$n\in\mathbb{N}$, we know that there exists a program index $a$ such
		that $a \sim_\pi^n h(a)$ holds. This means that the program
		transform $h$ does not alter the asymptotic complexity of, at least,
		the program $a$.  \qed
	\end{example}
	
	Our second recursion theorem allows us to obtain an intensional version of Rice's theorem for fair and ssmn abstract semantics.
	Inspired by~\cite{MS19}, we generalise the statement to cover partially extensional properties.
	
	% \begin{restatable}[Rice by fair and ssmn semantics]{theorem}{riceIntensional}
	\begin{theorem}[{\bf Rice by fair and ssmn semantics}]
		\label{th:rice-intensional}
		Let $\langle \pi, \equiv_\pi \rangle$ be a fair and ssmn semantics. If
		$A\subseteq \mathbb{N}$ is partially $\sim_{\pi}^n$-extensional and not universally
		$\sim_{\pi}^n$-extensional, for some arity $n\in \mathbb{N}$, then $A$ is not recursive.
	\end{theorem}
	\begin{proof}
		Since $A$ is partially $\sim_{\pi}^n$-extensional and not
		universally $\sim_{\pi}^n$-extensional, there are
		$x_0, x_1\in\mathbb{N}$ such that
		$[x_0]_{\sim_{\pi}^n}\cap A = \varnothing$ and
		$[x_1]_{\sim_{\pi}^n}\subseteq A$. Assume $A$ is recursive, hence
		its characteristic function $\chi_A$ is computable. Then, we can
		define a function $f:\mathbb{N}\rightarrow\mathbb{N}$ defined as
		follows:
		\[
		f(x)\eqd
		\begin{cases}
			x_0 & \text{ if }x\in A\\
			x_1& \text{ otherwise}
		\end{cases}\;\;=x_0\cdot\chi_A(x)+x_1\cdot (1- \chi_A(x)).
		\]
		Observe that $f$ is clearly total and computable. We can now apply our
		intensional second recursion Theorem~\ref{th:second-recursion}, and
		obtain that there exists $a\in \mathbb{N}$ such that
		$f(a) \sim_{\pi} a$. This easily leads to a contradiction that closes the proof. In fact, there are two cases, either $a \in A$
		or $a \notin A$.
		\begin{enumerate}
			\item If $a\in A$ then $f(a)=x_0 \sim_{\pi} a$ and thus, since $[x_0]_{\sim\pi}\cap A = \varnothing$, we have the contradiction $a \notin A$.
			\item Similarly, if $a \notin A$ then $f(a)=x_1 \sim_{\pi} a$ and
			thus, since $[x_1]_{\sim}\subseteq A$, we deduce the contradiction $a \in A$. \qedhere
		\end{enumerate}
	\end{proof}
	
	\begin{figure}[t]
		\centering{
			\begin{tikzpicture}[scale=1]
				\footnotesize
				\draw[thick, fill=gray!10] (0,0) ellipse [x radius=70pt, y radius =30pt];
				\draw[red,thick,fill=red!10] (-0.9,0) ellipse [x radius=32pt, y radius =18pt];
				\draw[red] (-0.1,0) node {$A$};
				\draw[blue,thick,fill=blue!10] (1.4,0) ellipse [x radius=18pt, y radius =13pt];
				\draw (1.4,0) node {$[a_0]_{\sim_\pi}$};
				\draw[blue,thick,fill=blue!10] (-1,0) ellipse [x radius=18pt, y radius =13pt];
				\draw (-1,0) node {$[a_1]_{\sim_\pi}$};
			\end{tikzpicture}
		}
		\caption{A graphical representation of Theorem~\ref{th:rice-intensional}.}
		\label{fig:second-rec}
	\end{figure}
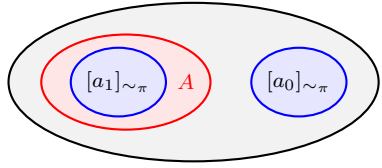

	Fig.~\ref{fig:second-rec} provides a graphical representation of this result:
	if we can find two program indices $a_0, a_1\in \mathbb{N}$ such that $A$
	over-approximates the $\equiv_\pi$-equivalence class $[a_1]_{\sim_\pi}$ and $A$ does not intersect $[a_0]_{\sim_\pi}$, then $A$ cannot be recursive.  Let us 
	illustrate some  applications of Theorem~\ref{th:rice-intensional}.

	\begin{example}[{\bf Halting set}]
		\label{ex:concrete-undec}
		Let $\langle \phi, \equiv_{\mathbb{N}} \rangle$ be the domain
		semantics of Example~\ref{ex:domain} with $S=\mathbb{N}$, hence
		$f \equiv_{\mathbb{N}} g$ when $\dom(f)=\dom(g)$.  The halting set
		$K\eqd \{a \in\mathbb{N} \mid \phi_a(a)\!\downarrow \}$ can be proved to be
		non-recursive by resorting to Theorem~\ref{th:rice-intensional} for $\langle \phi, \equiv_{\mathbb{N}} \rangle$. 
		Let $e_0,e_1\in\mathbb{N}$ be such that
		$\phi_{e_0}= \lambda x. \!\uparrow$ and $\phi_{e_1}=\lambda x.1$. Since
		$[e_1]_{\equiv_{\mathbb{N}}}$ is the set of programs that compute
		total functions, we have that $[e_1]_{\equiv_{\mathbb{N}}} \subseteq K$. Moreover, 
		$[e_0]_{\equiv_{\mathbb{N}}}$ is the set of nonterminating 
		programs for any input, so that
		$[e_0]_{\equiv_{\mathbb{N}}} \cap K=\varnothing$. This means that $\langle \phi, \equiv_{\mathbb{N}} \rangle$ satisfies the hypotheses of 
		Theorem~\ref{th:rice-intensional}, thus entailing that $K$ is not recursive.
		\qed
	\end{example}
	
	\begin{example}[{\bf Complexity sets}]
		\label{ex:complexity-undec}
		Let $\tuple{\phi,=}$, $\tuple{\Phi,\equiv_\Phi}$ be, resp., the
		semantics of Examples~\ref{ex:concrete} and~\ref{ex:complexity}. As
		observed in Section~\ref{se:smn-fair}, on a suitable computational
		model such as single tape Turing machines, these are fair ssmn
		semantics, so that Theorem~\ref{th:rice-intensional} applies.
		
		Let $\mathit{sort} : \mathbb{N} \to \mathbb{N}$ be a total function that
		takes as input an encoded sequence of numbers and 
		outputs the encoding of the corresponding sorted sequence. It turns out that by applying Theorem~\ref{th:rice-intensional}, the following
		sets can be proved to be non-recursive:
		
		\begin{enumerate}[(1)]
			\item $A\eqd \{a \mid \Phi_a\in \Theta(n \log n) \land \phi_a = \mathit{sort} \}$,
			\item $B\eqd\{a \mid \Phi_a\in \mathcal{O}(n \log n)\}$,
			\item $C\eqd\{a \mid \Phi_a\in \Omega(n \log n)\}$.
		\end{enumerate}
		
		Let $\textit{is}$, $\textit{ms}$ be different implementations of
		$\mathit{sort}$, i.e., $\phi_{\textit{is}}=\phi_{\textit{ms}}=sort$, such that 
		$\Phi_{\textit{is}}\in \Theta(n^2)$ and $\Phi_{\textit{ms}}\in \Theta(n\log n)$~---~$\textit{is}$ and $\textit{ms}$ could be, resp.,  insertion and merge sort.
		Recall that $\mathbin{\sim_R}$ denotes the Rice equivalence induced by $\tuple{\phi,=}$ (i.e., 
		$a\mathbin{\sim_R} b \Leftrightarrow \phi_a = \phi_b$), 
		and, in turn, let
		$\mathbin{\sim_{\Phi R}}=\mathbin{\sim_\Phi} \cap \mathbin{\sim_R}$ be the equivalence induced by
		the complexity clique semantics of Example~\ref{ex:asperti}, which is
		a fair ssmn semantics. Then, we have that:
		
		\begin{enumerate}[(1)]
			\item since $[\textit{is}\:\!]_{\sim_{\Phi R}}\cap A=\varnothing$ and $[\textit{ms}\:\!]_{\sim_{\Phi R}}\subseteq A$, by Theorem $\ref{th:rice-intensional}$, $A$ is non-recursive;
			\label{en:first}
			\item since $[\textit{is}\:\!]_{\sim_\Phi}\cap B=\varnothing$ and $[\textit{ms}\:\!]_{\sim_\Phi}\subseteq B$, by Theorem $\ref{th:rice-intensional}$, $B$ is non-recursive; \label{en:second}
			\item let $e$ be any program index such that $\Phi_{e}\in \Theta(1)$. Since $[e]_{\sim_{\Phi}}\cap C=\varnothing$ and $[\textit{is}\:\!]_{\sim_{\Phi}}\subseteq C$, by Theorem $\ref{th:rice-intensional}$, the set $C$ is non-recursive. \label{en:third} \qed
		\end{enumerate}
	\end{example}

	It is worth remarking that in Example~\ref{ex:complexity-undec},
	$n \log n$ could be replaced by any function, thus showing the undecidability of the
	asymptotic complexities ``big O'' (case~\eqref{en:second}) and 
	``big Omega'' (case~\eqref{en:third}). Let us also point out that Example~\ref{ex:concrete-undec} shows
	how easily the halting set $K$ can be proved to be 
	non-recursive by applying Theorem~\ref{th:rice-intensional}.
	
	\subsection{Branching Semantics}
	\label{se:branching-semantics}

	Let us investigate the connection between our results and the key
	notion of intricated switching family used by Moyen and Simonsen~\cite{MS19}
	for proving their intensional version of Rice's theorem. Firstly, we argue that every ssmn abstract semantics admits an intricated
	switching family whenever it is able to express a suitable form of
	\emph{conditional branching}. This allows us to derive an intensional Rice's theorem. Moreover, we show that for fair and ssmn semantics, the identity can always play
	the role of intricated switching family.
	
	\begin{definition}[{\bf Branching and discharging semantics}]
		\label{de:branching}
		An abstract semantics $\langle \pi, \equiv_\pi\rangle$ is \emph{branching} if, given $n\geq 1$, there exists a total computable function
		$r:\mathbb{N}^4 \to \mathbb{N}$ such that $\forall a,b,c_1,c_2,x\in\mathbb{N}$ with $c_1\neq c_2$:
		\begin{equation*}
			\lambda \Vec{y}.\pi^{(n)}_{r(a,b,c_1,c_2)}(x,\Vec{y})\equiv_\pi
			\begin{cases}
				\lambda \Vec{y}.\pi_a^{(n)}(x,\Vec{y}) & \text{ if } x=c_1\\
				\lambda \Vec{y}.\pi_b^{(n)}(x,\Vec{y}) & \text{ if } x=c_2\\
				\lambda \Vec{y}.\!\uparrow & \text{ otherwise. } 
			\end{cases} 
		\end{equation*}
		Moreover, $\langle \pi, \equiv_\pi\rangle$ is 
		(variable) \emph{discharging} if, for all $n\geq 1$, there exists a total computable function $t:\mathbb{N}\to\mathbb{N}$  such that for all $a,x\in\mathbb{N}$:
		\begin{equation*}
			\pi^{(n)}_{a} \equiv_\pi \lambda \Vec{y}.\pi_{t(a)}^{(n+1)}(x,\Vec{y}). 
			\tag*{\qed}
		\end{equation*}
	\end{definition}
	
	Hence, intuitively, an abstract semantics is branching when 
	it is able to model the branching structure of conditional statements 
	with multiple positive guards, while the property of being 
	variable discharging holds when 
	one can freely add fresh and unused 
	variables without altering the abstract semantics. Let us first recall the notion of recursive
	inseparability~\cite[Section 3]{sr58} and of intricated switching family from~\cite[Definition~5]{MS19}.\footnote{For the sake of simplicity, \cite[Definition~5]{MS19} is 
		here instantiated to the case of recursive sets.}
	
	\begin{definition}[{\bf Recursively inseparable sets}]
		\label{def:recursivelyinseparable}
		Two sets $A,B\subseteq\mathbb{N}$ of program indices
		are \emph{recursively inseparable} if there exists no 
		$C\subseteq\mathbb{N}$ such that
		$A\subseteq C$ and $B\cap C=\varnothing$. \qed
	\end{definition}
	
	\begin{definition}[{\bf Intricated switching family {\cite[Definition~5]{MS19}}}]
		\label{de:switching-family}
		Let $\mathbin{\sim}\subseteq\mathbb{N}\times\mathbb{N}$ be an equivalence
		relation on program indices. An \emph{intricated switching family} (ISF) w.r.t.\ 
		$\sim$ is an indexed set of total computable functions
		$\{ \sigma_{a,b} \}_{a,b \in \mathbb{N}}$, with
		$\sigma_{a,b} : \mathbb{N} \to \mathbb{N}$, such that for all
		$a, b \in \mathbb{N}$, the sets
		$A_{a,b}=\{x\in \mathbb{N}\mid\sigma_{a,b}(x)\sim a\}$ and
		$B_{a,b}=\{x\in \mathbb{N}\mid\sigma_{a,b}(x)\sim b\}$ are recursively
		inseparable.\qed
	\end{definition}
	
	Moyen and Simonsen~\cite[Theorem~3]{MS19} show that if 
	an equivalence $\sim$ admits an ISF,  then every partially $\sim$-extensional and not universally $\sim$-extensional set is not recursive. A simplified version of their intensional result, tailored for our setting, can be stated as follows.
	
	\begin{theorem}[{\bf {\cite[Theorem 3]{MS19}}}]
		\label{th:ms19}
		Let
		$\mathbin{\sim}\subseteq\mathbb{N}\times\mathbb{N}$ be an
		equivalence relation.  If $A\subseteq \mathbb{N}$ is partially
		$\sim$-extensional, not universally $\sim$-extensional and there
		exists an ISF w.r.t.\ $\sim$ then $A$ is not recursive.
	\end{theorem}
	
	Branching discharging and ssmn semantics can be shown to admit an
	intricated switching family, in a way that, relying on
	Theorem~\ref{th:ms19} we can derive the following intensional
	version of Rice's Theorem.
	
	% \begin{restatable}[Rice by branching, discharging and ssmn semantics]{theorem}{riceBranching}
	\begin{theorem}[{\bf Rice by branching, discharging and ssmn semantics}]
		\label{th:rice-branching}
		Let $\langle \pi, \equiv_\pi\rangle$ be a branching, discharging and ssmn semantics. If $A\subseteq \mathbb{N}$ is partially
		$\sim_{\pi}^n$-extensional and not universally $\sim_{\pi}^n$-extensional for some arity $n\in \mathbb{N}$,
		then $A$ is not recursive.
	\end{theorem}
	
	\begin{proof}
		Let $u\in\mathbb{N}$ be an index for the standard
		unary universal program.  Consider the total computable
		functions $r:\mathbb{N}^4\to\mathbb{N}$ and
		$t:\mathbb{N}\to\mathbb{N}$ of, resp., the branching and variable
		discharging properties. By ssmn property, there exists a total
		computable function $s:\mathbb{N}^4\rightarrow\mathbb{N}$ such that
		$\forall a,b,x\in\mathbb{N}$:
		\begin{align*}
			\pi_{s(r(t(a),t(b),0,1),u,x,0)}^{(n)}&\equiv_\pi  \lambda \Vec{y}.\pi_{r(t(a),t(b),0,1)}^{(n+1)}(\phi_u^{(2)}(x,0), \Vec{y}) \tag*{[by ssmn property]}\\
			&=\lambda \Vec{y}.\pi_{r(t(a),t(b),0,1)}^{(n+1)}(\phi_x(0), \Vec{y})\\
			&\equiv_\pi
			\begin{cases}
				\lambda \Vec{y}.\pi_{t(a)}^{(n+1)}(0,\Vec{y}) & \text{ if } \phi_x(0)=0 \\
				\lambda \Vec{y}.\pi_{t(b)}^{(n+1)}(1,\Vec{y}) & \text{ if } \phi_x(0)=1\\
				\lambda \Vec{y}.\uparrow & \text{ otherwise}
			\end{cases} \tag*{[by branching property]}
			\\
			&\equiv_\pi
			\begin{cases}
				\pi_{a}^{(n)} & \text{ if } \phi_x(0)=0 \\
				\pi_{b}^{(n)} & \text{ if } \phi_x(0)=1 \\
				\lambda \Vec{y}.\uparrow & \text{ otherwise}
			\end{cases} \tag*{[by variable discharging property]}
		\end{align*}

		For all $a,b\in\mathbb{N}$, we define the total computable function
		$$\sigma_{a,b}(x) \eqd s(r(t(a),t(b),0,1),u,x,0).$$ We claim that the
		family of functions $\{\sigma_{a,b}\}_{a,b\in \mathbb{N}}$ is
		intricated w.r.t. $\sim_{\pi}^n$ (cf.\
		Definition~\ref{de:switching-family}).
		In fact, for all  $a,b \in \mathbb{N}$, let $A_{a,b}\eqd \{x\in\mathbb{N} \mid \sigma_{a,b}(x)\sim_{\pi}^n a\}$ and  $B_{a,b}\eqd \{x\in\mathbb{N} \mid \sigma_{a,b}(x)\sim_{\pi}^n
		b\}$.
		We have four cases:
		\begin{enumerate}
			
			\item if $\pi_a^{(n)} \equiv_\pi \pi_b^{(n)}$, then $A_{a,b}=B_{a,b}$ and therefore they are trivially recursively inseparable;
			
			\item if $\pi_a^{(n)} \not\equiv_\pi \pi_b^{(n)}$ and $\pi_a^{(n)}\not\equiv_\pi \lambda \Vec{x}.\uparrow\not\equiv_\pi \pi_b^{(n)}$ we have $A_{a,b}=\{x\in\mathbb{N}\mid \phi_x(0)=0\}$ and $B_{a,b}=\{x\in\mathbb{N}\mid \phi_x(0)=1\}$. Hence, the sets $A_{a,b}$ and $B_{a,b}$ are recursively inseparable (cf.\ \cite[Section 3.3]{papadimitriou94});
			\item if $\pi_b^{(n)}\not\equiv_\pi \pi_a^{(n)}\equiv_\pi\lambda \Vec{x}.\!\uparrow$ we have $B_{a,b}=\{x\in\mathbb{N}\mid \phi_x(0)=1\}$ and $A_{a,b} = \{x\in\mathbb{N}\mid \phi_x(0) \neq 1 \} = \overline{B_{a,b}}$. The mere fact that $B_{a,b}$ is not recursive (by classical Rice's Theorem) thus implies that $A_{a,b}$ and $B_{a,b}$ are not recursively separated;
			%\item if $\pi_b^{(n)}\not\equiv_\pi \pi_a^{(n)}\equiv_\pi\lambda \Vec{x}.\!\uparrow$ we have $\{x\in\mathbb{N}\mid \phi_x(0)%=0\}\subseteq \{x\in\mathbb{N}\mid \phi_x(0)=0 \lor {\phi_x(0)\!\uparrow\}}=A_{a,b}$ and $B_{a,b}=\{x\in\mathbb{N}\mid %\phi_x(0)=1\}$. Assume that $A_{a,b}$ and $B_{a,b}$ are recursively separated by a set $C$: since $\{x\in\mathbb{N}\mid %\phi_x(0)=0\}\subseteq A_{a,b}\subseteq C$, this leads to the contradiction that $\{x\in\mathbb{N}\mid \phi_x(0)=0\}$ and %$B_{a,b}$ are recursively separated by $C$;
			%    \todo[inline]{P: Mi pare si possa provare piu' direttamente, mi sugge qualcosa?\\
			%      if $\pi_b^{(n)}\not\equiv_\pi \pi_a^{(n)}\equiv_\pi\lambda \Vec{x}.\!\uparrow$ we have $B_{a,b}=\{x\in\mathbb{N}\mid \phi_x(0)=1\}$ and $A_{a,b} = \{x\in\mathbb{N}\mid \phi_x(0) \neq 1 \} = \overline{B_{a,b}}$. The mere fact that $B_{a,b}$ is not recursive thus implies that $A_{a,b}$ and $B_{a,b}$ are not recursively separated;\\
			%	  L: sì, ho adattato la dimostrazione dei casi 3 e 4.
			%  }
			\item if $\pi_a^{(n)}\not\equiv_\pi \pi_b^{(n)}\equiv_\pi\lambda \Vec{x}.\!\uparrow$ we can take $a'=b$ and $b'=a$ and conclude by Case 3.
			%\item if $\pi_a^{(n)}\not\equiv_\pi \pi_b^{(n)}\equiv_\pi\lambda \Vec{x}.\!\uparrow$ we have $A_{a,b}=\{x\in\mathbb{N}\mid \phi_x(0)=0\}$ and $\%{x\in\mathbb{N}\mid \phi_x(0)=1\}\subseteq B_{a,b}=\{x\in\mathbb{N}\mid\phi_x(0)=1 \lor \phi_x(0)\uparrow\}$. Again, if $A_{a,b}$ and $B_{a,b}$ %were recursively separated by a set $C$ we would have $\{x\in\mathbb{N}\mid \phi_x(0)=1\}\cap C=\varnothing$, leading to the contradiction that $A_%{a,b}$ and $\{x\in\mathbb{N}\mid \phi_x(0)=1\}$ are recursively separated by $C$.
		\end{enumerate}
		Since in all cases $A_{a,b}$ and $B_{a,b}$ are recursively inseparable, it turns out that $\{\sigma_{a,b}\}_{a,b\in\mathbb{N}}$ is an ISF w.r.t.\ $\sim_{\pi}^n$ and thus we conclude by Theorem \ref{th:ms19}.
	\end{proof}
	
	Let us discuss more in detail the relationship with the approach in \cite{MS19}. Firstly, let us show a lemma which will be fundamental to prove the following results.
	
	\begin{lemma}
		\label{lemma:id-intrication}
		Let $\sim$ be an equivalence relation on program indices. If every set
		$A$ partially $\sim$-extensional and not universally $\sim$-extensional is non-recursive then the identity {\rm$\Id$} is an ISF w.r.t.\ $\sim$.
	\end{lemma}
	\begin{proof}
		Clearly, the identity $\Id \eqd\{(\lambda x.x)_{a,b}\}_{a,b\in\mathbb{N}}$ is a family of total computable functions. Moreover, for
		$a, b \in \mathbb{N}$ we have
		$A_{a,b}=\{x\in\mathbb{N}:x\sim a\}=[a]_\sim$ and
		$B_{a,b}=\{x\in\mathbb{N}:x\sim b\}=[b]_\sim$. Therefore, every
		set $C \subseteq \mathbb{N}$ such that $A_{a,b} \subseteq C$ and
		$B_{a,b} \cap C = \varnothing$, is partially $\sim$-extensional and not
		universally $\sim$-extensional and thus, by hypothesis, not
		recursive. Hence, $A_{a,b}$ and $B_{a,b}$ are recursively
		inseparable.
	\end{proof}
	
	It turns out that a fair ssmn semantics always admits  
	a canonical ISF, namely, the identity $\Id \eqd\{(\lambda x.x)_{a,b}\}_{a,b\in\mathbb{N}}$.  
	
	% \begin{restatable}{proposition}{identityIsIntricated}
	\begin{proposition}
		Let $\langle \pi, \equiv_\pi\rangle$ be a fair and ssmn semantics.
		Then, the identity {\rm $\Id$} is an ISF w.r.t.\ $\sim_\pi^n$, for
		all $n\geq 1$.
	\end{proposition}
	
	\begin{proof}
		Since $\langle \pi, \equiv_\pi\rangle$ is a fair ssmn semantics, by
		Theorem~\ref{th:rice-intensional}, every partially
		$\sim_\pi^n$-extensional and not universally
		$\sim_\pi^n$-extensional set $A$ is non-recursive. Therefore, we
		conclude by applying Lemma~\ref{lemma:id-intrication}.
	\end{proof}
	
	Let us point out that the identity function has not been exploited in~\cite{MS19}, that instead focuses on the standard switching family. It turns out that the identity function plays a key role as ISF.

	% \begin{restatable}{proposition}{identityMostGeneral}
	\begin{proposition}
		\label{prop:identity}
		Let $\mathbin{\sim}\subseteq\mathbb{N}\times\mathbb{N}$ be an
		equivalence relation. The following statements are equivalent:
		\begin{enumerate}[(1)]
			\item Every set $A\subseteq \mathbb{N}$ partially $\sim$-extensional and not universally $\sim$-extensional is non-recursive.
			\item The identity {\rm$\Id$} is an ISF  w.r.t.\ $\sim$.
			\item There exists an ISF w.r.t.\ $\sim$.
		\end{enumerate}
	\end{proposition}
	
	\begin{proof}\ \
		\begin{itemize}
			\item[] $(1\Rightarrow 2)$: by Lemma~\ref{lemma:id-intrication};
			\item[] $(2\Rightarrow 3)$: trivial;
			\item[] $(3\Rightarrow 1)$: by Theorem~\ref{th:ms19}. \qedhere
		\end{itemize}
	\end{proof}
	
	%\medskip
	Therefore, the above result roughly states that the identity
	function is the ``canonical'' ISF, meaning that if an ISF exists, then $\Id$ 
	is an ISF 
	as well.
	Moreover, the intensional Rice's Theorem~\ref{th:ms19} of
	\cite{MS19} provides a sufficient condition (i.e., the existence of an
	ISF) for a partially and not universally
	extensional set to be undecidable. Proposition~\ref{prop:identity} 
	enhances Theorem~\ref{th:ms19} by
	showing that such a sufficient condition is necessary as well, or, equivalently, that a
	partially and not universally extensional set is undecidable iff 
	there exists an ISF. 
	
	We conclude this section by discussing an alternative notion of branching, which requires the preservation of a full conditional statement with positive and negative guards.
	
	\begin{definition}[{\bf Strongly branching semantics}]
		\label{de:strongly-branching}
		An abstract semantics ${\langle \pi, \equiv_\pi\rangle}$ is \emph{strongly branching}
		if, given $n\geq 1$, there exists a total computable function
		$r : \mathbb{N}^3 \to \mathbb{N}$ such that
		for all $a, b,c,x\in \mathbb{N}$:
		\[\lambda \Vec{y}.\pi^{(n)}_{r(a,b,c)}(x, \Vec{y}) \equiv_\pi
		\begin{cases}
			\lambda \Vec{y}.\pi_a^{(n)}(x,\Vec{y}) & \text{ if } x=c\\
			\lambda \Vec{y}.\pi_b^{(n)}(x,\Vec{y}) & \text{ otherwise. }
		\end{cases}\tag*{\qed}
		\] 
	\end{definition}
	
	The condition above is an adaptation to our framework of a property that is needed in order to exploit a  so-called standard
	switching family as defined in~\cite[Example~1]{MS19}.
	Despite appearing to be more natural, the preservation of conditionals with
	positive and negative conditions is a stronger requirement than the
	one we considered in Definition~\ref{de:branching}. Indeed, it turns
	out that every ssmn and strongly branching semantics is a branching
	semantics.
	
	% \begin{restatable}[Strongly branching implies branching]{proposition}{stronglyBranching}
	\begin{proposition}[{\bf Strongly branching implies branching}]
		\label{pr:strongly-branching}
		If $\langle \pi,\equiv_\pi\rangle$ is a ssmn and strongly branching
		semantics, then $\langle \pi,\equiv_\pi\rangle$ is a branching semantics.
	\end{proposition}
	
	\begin{proof}
		Given an arity $n$, let $r$ be the function of the strongly branching property
		of Definition~\ref{de:strongly-branching}. By \eqref{prop:smn-undef-function} there exists an index $e_0\in\mathbb{N}$ such that $\pi_{e_0}^{(n)} \equiv_{\pi} \lambda \Vec{y}.\!\uparrow$.
		Now, we define the function $\sigma:\mathbb{N}^4\to\mathbb{N}$ such that for all $a,b,c_1,c_2 \in \mathbb{N}$ we have $\sigma(a,b,c_1,c_2)=r(a, r(b,e_0,c_2),c_1)$. Note that $\sigma$ is a total computable function, by composition, and for all $a,b, c_1, c_2,x\in\mathbb{N}$ with $c_1 \neq c_2$:
		\begin{align*}
			\lambda \Vec{y}.\pi^{(n)}_{\sigma(a,b,c_1,c_2)}(x,\Vec{y})
			& =
			\lambda \Vec{y}.\pi^{(n)}_{r(a, r(b,e_0,c_2),c_1)}(x,\Vec{y})\\
			&\equiv_\pi
			\begin{cases}
				\lambda \Vec{y}.\pi_{a}^{(n)}(x,\Vec{y}) & \text{ if } x=c_1\\
				\lambda \Vec{y}.\pi_{r(b,e_0,c_2)}^{(n)}(x,\Vec{y}) & \text{ otherwise }
			\end{cases}
			\tag*{[by branching property]}
			\\
			&\equiv_\pi
			\begin{cases}
				\lambda\Vec{y}.\pi_{a}^{(n)}(x,\Vec{y}) & \text{ if } x=c_1\\
				\lambda\Vec{y}.\pi_{b}^{(n)}(x,\Vec{y}) & \text{ if } x\neq c_1 \land x=c_2\\
				\lambda\Vec{y}.\pi_{e_0}^{(n)}(x,\Vec{y}) & \text{ if } x\neq c_1 \land x\neq c_2\\
			\end{cases}
			\tag*{[by branching property]}
			\\
			&\equiv_\pi
			\begin{cases}
				\lambda\Vec{y}.\pi_a^{(n)}(x,\Vec{y})  & \text{ if } x=c_1\\
				\lambda\Vec{y}.\pi_b^{(n)}(x,\Vec{y}) & \text{ if } x=c_2\\
				\lambda \Vec{y}.\uparrow & \text{ otherwise }
			\end{cases}\\
		\end{align*} 
		Thus, $\sigma$ is the desired function for the branching property.
	\end{proof}
	
	\subsection{An Application to Static Program Verifiers}
	\label{ss:examples-verifier}
	We adapt the general definition of static program verifier of Cousot et al.~\cite[Definition~4.3]{CGR18} to our framework. Given a program 
	property $P \subseteq \mathbb{N}$ to check, a static program verifier is a total recursive function $\mathcal{V}:\mathbb{N}\rightarrow \{0,1\}$. It is 
	\emph{sound} when for all $p\in\mathbb{N}$, $\mathcal{V}(p)=1 \Rightarrow p \in P$, while $\mathcal{V}$ is \emph{precise} if the reverse implication also holds, i.e., when $\mathcal{V}(p)=1\Leftrightarrow p \in P$ holds.
	Informally, soundness guarantees that only false negatives are allowed, i.e., $\mathbb{N}\smallsetminus P$ is possibly a proper subset of $\{p\in\mathbb{N}:\mathcal{V}(p)=0\}$, while precise verifiers output true positives and true negatives only (i.e., they decide $P$).
	
	Classical Rice's theorem clearly entails the impossibility of designing 
	a precise verifier for a nontrivial extensional property. However, one may wonder whether there exist sound verifiers with ``few'' false negatives. By
	applying our intensional Theorem~\ref{th:rice-intensional}, we are able to show that sound but imprecise verifiers necessarily have at least one false negative for each equivalence class of programs, even for intensional properties.
	
	\begin{example}[{\bf Constant value verifier}]
		\label{ex:verifier-division}
		Assume we are interested in checking if a program can output a given
		constant value, for instance, zero with the aim of statically
		detecting division-by-zero bugs. Let $\mathcal{V}$ be a sound static
		verifier for the set
		$P_{=0}\eqd \{p \in \mathbb{N}\mid 0\in \rng(\phi_p)\}$ of programs
		that output zero for some input. The set
		$N \eqd \{p \in \mathbb{N} \mid \mathcal{V}(p)=0 \}$ is recursive
		since $\mathcal{V}$ is assumed to be a total computable function. By
		soundness of $\mathcal{V}$, we have that
		$\mathbb{N}\smallsetminus P_{=0} \subseteq N$, so that $N$ includes,
		for example, the set of programs computing the constant function
		$\lambda x.1$. Therefore, $N$ is partially extensional, and, by
		Theorem~\ref{th:rice-intensional}, $N$ has to be universally
		extensional. This means that for any computable function $f\in \mathcal{C}$ there
		exists a program $p \in \mathbb{N}$ that computes $f$ such that
		$\mathcal{V}(p)=0$. Thus, when $0\in \rng(f)$ holds (e.g.,
		for $f=\lambda x.0$), $\mathcal{V}$ necessarily outputs a false
		negative for $p$.  Hence, $\mathcal{V}$ outputs infinitely many
		false negatives.  \qed
	\end{example}
	
	\begin{example}[{\bf Complexity verifier}]
		\label{ex:verifier-complexity}
		Consider a speculative sound static verifier $\mathcal{V}$ for
		recognizing programs that meet some lower bound, for instance,
		programs having a cubic lower bound
		$P_{\Omega(n^3)}\eqd \{ p \in \mathbb{N} \mid
		\Phi_p=\Omega(n^3)\}$. Thus,
		$N \eqd \{ p \in \mathbb{N} \mid \mathcal{V}(p)=0\}$ has to be
		recursive and if $\sim_\Phi$ is the program equivalence induced by
		the Blum complexity semantics $\langle \Phi, \equiv_\Phi\rangle$ of
		Example~\ref{ex:complexity} then, by soundness of $\mathcal{V}$, we
		have, for example,
		$\{p\in\mathbb{N} \mid \Phi_p=\Theta(1)\}\subseteq N$. This means
		that $N$ is partially $\sim_\Phi$-extensional and, by
		Theorem~\ref{th:rice-intensional}, $N$ is universally extensional,
		namely, $\mathcal{V}$ will output 0 for at least a program in each
		Blum complexity class. For instance, even some programs with an
		exponential lower bound will be wrongly classified by $\mathcal{V}$
		as programs that do not meet a cubic lower bound.  \qed
	\end{example}
	
	As shown by Cousot et al.~\cite[Theorem~5.4]{CGR18}, precise static verifiers cannot be designed
	(unless for trivial program properties).
	The examples above prove that, additionally, we cannot have any certain information on an input program $p$ whenever the output of a sound (and imprecise) verifier for $p$ is $0$. In fact, when this happens, $p$ could compute any partial function (cf.~Example~\ref{ex:verifier-division}) or have any complexity (cf.~Example~\ref{ex:verifier-complexity}).

	\section{On the Decidability of Affine Program Invariants}
	\label{se:cfg}
	
	Karr's abstract domain~\cite{karr76} 
	consisting of affine equalities between program variables,
	such as $2x-3y=1$, is well known and widely used in 
	static program analysis~\cite{cousot21,mine17,RY20}. 
	Karr~\cite{karr76} put forward an algorithm that infers for each program point $q$ of a control flow 
	graph modelling an affine program $P$ (i.e., an unguarded program with non-deterministic branching and affine assignments) a set of affine 
	equalities that hold among the variables of $P$ when the control reaches $q$, namely, an \emph{affine invariant} for $P$.  
	M\"uller-Olm and Seidl~\cite{MOS04Affine} show that Karr's algorithm actually 
	computes the strongest affine invariant for affine programs (this result has been extended to a slightly larger class of affine programs 
	in \cite[Theorem~5.1]{ran20}). Moreover, they design 
	a more efficient algorithm implementing this static analysis and they
	extend in~\cite{MOS04Poly}  this algorithm for computing bounded polynomial invariants, i.e., 
	the strongest polynomial equalities of degree at most a given $d\in \mathbb{N}$. 
	Later, Hrushovski et al.~\cite{Worrel18} put forward a sophisticated
	algorithm for computing the strongest unbounded 
	polynomial invariants of affine programs, by relying on the Zariski
	closure of semigroups. 
	\\
	\indent
	On the impossibility side, M\"uller-Olm and Seidl~\cite[Section~7]{MOS04Affine} 
	prove that
	for affine programs allowing positive affine  
	guards it is undecidable 
	whether a given nontrivial affine equality holds at a given program point or not. 
	In practical applications, static 
	analyses on Karr's abstract domain of guarded affine programs ignore non-affine Boolean
	guards, while for an
	affine guard $b$, the current affine invariant $i$ is propagated through 
	the positive branch of $b$ by the intersection $i\cap b$, that remains an affine subspace. 
	By the aforementioned undecidability result~\cite[Section~7]{MOS04Affine}, 
	this latter analysis algorithm for guarded affine programs turns out to be sound but necessarily imprecise, thus inferring affine invariants that, in general, might not be the strongest ones.   
	M\"uller-Olm and Seidl~\cite[Section~7]{MOS04Affine} prove
	their undecidability result by exploiting an acute reduction
	to the undecidable Post correspondence problem, inspired by early 
	reductions explored in data flow analysis \cite{hecht,ku77}. 
	In this section, we show that our Theorem~\ref{th:rice-branching} allows us to derive and extend this
	undecidability result by exploiting an orthogonal intensional approach. 
	More precisely, we prove that any nontrivial (and not necessarily affine) relation 
	on the states of control flow graphs of programs allowing: (1)~zero, variable and 
	successor assignments, resp., $x:=0$, $x:=y$ and $x:=y+1$, and
	(2)~positive 
	equality guards $x=y?$ and $x=v?$, turns out to be 
	undecidable. Since these control flow graphs form a subclass of  
	affine programs with positive affine guards, 
	the undecidability result of M\"uller-Olm and Seidl~\cite[Section~7]{MOS04Affine} 
	is retrieved as a consequence.

	Following the standard approach, we consider control flow graphs that consist of program points connected by edges labeled by assignments and guards. Variables are denoted by $x_i$, with $i \in \mathbb{N}$, and store
	values ranging in $\mathbb{N}$, while Karr's abstract domain is  designed for variables assuming values in $\mathbb{Q}$. Clearly, from a computability perspective, this is not a restriction since one can 
	consider a computable bijection between $\mathbb{N}$ and $\mathbb{Q}$.
	
	\begin{definition}[{\bf Basic affine control flow graph}]\label{def:bacfg}
		A \emph{basic affine control flow graph} (BACFG) is a tuple $G=(N,E,s,e)$, where $N$ is
		a finite set of nodes, $s,e\in N$ are the start and end
		nodes, and $E \subseteq N \times \com \times N$ is a set
		of labelled edges, and the set $\com$ of commands consists of assignments of type
		$x_n:=0$, $x_n:=x_m$, $x_n:=x_m+1$,   and equality guards 
		of type $x_n=x_m?$, $x_n=v?$, with $v\in \mathbb{N}$. \qed
	\end{definition} 
	
	Let us remark that BACFGs only include basic affine assignments and 
	positive affine guards, in particular inequality checks such as $x_n\neq x_m?$ and $x_n\neq v?$ 
	are not allowed.  Thus, BACFGs are a subclass of
	affine programs with positive affine guards considered in \cite[Section~7]{MOS04Affine}.
	\\
	\indent
	As in dataflow analysis and abstract interpretation~\cite{cousot21,CC77,hecht,RY20}, 
	BACFGs have a \emph{collecting semantics}
	where, given a set of input states $\mathit{In}$, 
	each program point is associated with the set of 
	states that occur in some program execution from some state in $\mathit{In}$. 
	A finite number of variables may occur in a BACFG, so that a state of a BACFG $G$ is 
	a tuple $(x_1,\dots,x_k)\in\mathbb{N}^k$, where $k$ is the maximum variable index occuring in $G$ and $k=0$ is a degenerate case for trivial BACFGs with  
	$\mathbb{N}^0=\{\bullet\}$. 
	The \emph{collecting transfer function} $f_{(\cdot)}(\cdot):\com\to \wp(\mathbb{N}^k)\to \wp(\mathbb{N}^k)$ for $k\in \mathbb{N}$ variables and with $n,m\in [1,k]$ is 
	defined as  follows:
	\[
	\begin{split}
		f_{x_n:=0}(S) &\triangleq \{(x_1,\dots,x_{n-1}, 0,x_{n+1},\dots,x_k)\mid \Vec{x}\in S \},\\
		f_{x_n:=x_m}(S) &\triangleq \{(x_1,\dots,x_{n-1}, x_m,x_{n+1},\dots,x_k)\mid \Vec{x}\in S \},\\
		f_{x_n:=x_m+1}(S) &\triangleq \{(x_1,\dots,x_{n-1}, x_m+1,x_{n+1},\dots,x_k)\mid\Vec{x}\in S \},\\
		f_{x_n=v?}(S) &\triangleq \{\Vec{x}\in S \mid x_n=v \}, \\
		f_{x_n=x_m?}(S)& \triangleq \{\Vec{x}\in S \mid x_n=x_m \}.
	\end{split}
	\]
	A no-op  command denoted by $\epsilon$ is syntactic sugar for  $x_1:=x_1$, 
	i.e., $f_\epsilon  \triangleq f_{x_1:=x_1}=\lambda S.S$.
	Given $k,k'\in \mathbb{N}$ and 
	$S\in \wp(\mathbb{N}^{k'})$, 
	the projection 
	$S\!\restriction_{k}\in \wp(\mathbb{N}^{k})$ is defined as follows:
	\[
	S\!\restriction_{k}\:\eqd
	\begin{cases}
		S\times \mathbb{N}^{k-{k'}} & \text{ if } 0 \leq k' < k\\
		S& \text{ if } k'=k\\
		\{(x_1,\dots,x_k)\mid \Vec{x}\in S\}& \text{ if } k<k'\\
	\end{cases}
	\]
	
	\begin{definition}[{\bf Collecting semantics of BACFGs}]
		\label{de:collecting-semantics}
		Given a BACFG $G=(N,E,s,e)$ with $k\in \mathbb{N}$ variables and a set of input states 
		$S\subseteq \mathbb{N}^{k'}$, with $k'\leq k$, 
		the \emph{collecting semantics}
		${\llbracket G \rrbracket_{S}:N\rightarrow\wp(\mathbb{N}^k)}$ is the
		least, w.r.t.\ pointwise set inclusion, 
		solution in $\wp(\mathbb{N}^k)^{|N|}$
		of the following system of constraints: 
		\begin{equation*}
			\begin{cases}
				\llbracket G \rrbracket_{S} [s]\supseteq S\!\restriction_{k} &\text{ for the start node } s\\
				\llbracket G \rrbracket_{S} [v]\supseteq  f_c(\llbracket G \rrbracket_S [u]) &\text{ for each edge } (u,c,v)\in E
			\end{cases}\tag*{\qed}
		\end{equation*}
	\end{definition}
	
	Let us observe that, since the collecting 
	transfer functions $f_c$ are additive on the complete lattice $\tuple{\wp(\mathbb{N}^k),\subseteq}$,  by Knaster-Tarski fixpoint theorem, 
	$\grasse{G}_S$ is well defined. 
	For $\vec{x}\in\mathbb{N}^{k'}$, we write $\llbracket G \rrbracket _ {\vec{x}}$ instead of $\llbracket G \rrbracket_{\{\vec{x}\}}$.
	Notice that $\grasse{G}_{(\cdot)}$ is an additive function, so that, for any program point $u\in N$,  
	$\grasse{G}_S[u] = \bigcup_{\Vec{x}\in S} \grasse{G}_{\Vec{x}} [u]$ holds.

	\subsection{Turing Completeness of BACFGs}
	\label{subse:turing-completeness}
	
	Let us recall that a ssmn abstract semantics needs an underlying
	Turing complete concrete semantics of programs (cf.\
	Assumption~\ref{ass-tc}).  A crucial observation is that BACFGs are
	Turing complete despite not including full (both positive and
	negative) Boolean tests. This is proved by showing that any program of
	an Unlimited Register Machine (URM), which is a well-known Turing complete computational
	model~\cite{Cut80}, can be simulated by a BACFG.
	
	\begin{restatable}[{\bf Turing completeness of BACFGs}]{theorem}{cfgTuringCompleteness}
		\label{th:bacfg}
		\mbox{\rm BACFGs} are a  Turing complete computational model. 
	\end{restatable}

	Before getting into the technical details, it is worth providing, first, an
	intuition of the proof of Theorem~\ref{th:bacfg}. 
	Using the
	definition and notation of Cutland~\cite[Section 1.2]{Cut80}, let us recall 
	the four types of instructions of URMs:  
	
	\begin{itemize}
		\item $z(n)$: sets register $r_n$ to 0 ($r_n\leftarrow 0$) and transfers the control to the next instruction;
		\item $s(n)$: increments register $r_n$ by 1 ($r_n\leftarrow r_n+1$) and transfers the control to the next instruction;
		\item $t(m,n)$: sets register $r_n$ to $r_m$ ($r_n\leftarrow r_m$) and transfers the control to the next instruction;
		\item $j(m,n,p)$: if $r_m=r_n$ and $I_p$ is a proper instruction, then it jumps to the instruction $I_p$; otherwise, it skips to the next instruction;
	\end{itemize}
	
	It turns out that all these URM instructions 
	can be simulated by the BACFGs depicted in
	Figures~\ref{fig:turing-completeness-instr} and
	\ref{fig:turing-completeness-jump}. While the BACFGs in
	Figure~\ref{fig:turing-completeness-instr} are trivial, let us
	describe more in detail how the BACFG in
	Figure~\ref{fig:turing-completeness-jump} simulates a jump
	instruction $j(m,n,p)$. Intuitively, a difficulty arises for simulating the
	negative branch $x_n\neq x_m?$. Here, the BACFG at node $q_i$
	initialises a fresh unused variable $z$ with both $x_n+1$ and $x_m+1$
	and transfers the control to a node $inc_i$ where $z$ is incremented
	infinitely many times. Thus, in the least fixpoint solution, at node
	$inc_i$ the variable $z$ stores any value $v>\min(x_m, x_n)$,
	including $z=\max(x_m, x_n)$. Suppose now that $x_n>x_m$ holds: in
	this case, the guard $x_n=z?$ between nodes $inc_i$ and $q_{i+1}$
	eventually will be made true and at the node $q_{i+1}$ the store will
	retain the original values of all variables ($x_m$ and $x_n$
	included), except for the new variable $z$ which will be ignored by
	the remaining nodes. The case $x_m>x_n$ is analogous. Therefore, it
	turns out that the node $q_{i+1}$ will be reached if and only if
	$x_m \neq x_n$ holds, while $q_{p}$ will be reached if and only if
	$x_m = x_n$ holds, thus providing a simulation for the jump
	instruction $j(m,n,p)$.
	
	\begin{figure}
		\centering
		\begin{tikzpicture}[shorten >=0pt,node distance=2cm,on grid,>=stealth',every state/.style={inner sep=0pt, minimum size=7mm, draw=blue!50,very thick,fill=blue!10}]
			\node[state]         (q1)	{{$q_{i}$}}; 
			\node[state]         (q2) [below=of q1] {{$q_{i+1}$}};
			\path[->] 
			(q1) edge [right] node  {{$x_n:=0$}} (q2) ;
		\end{tikzpicture}
		\qquad\qquad
		\begin{tikzpicture}[shorten >=0pt,node distance=2cm,on grid,>=stealth',every state/.style={inner sep=0pt, minimum size=7mm, draw=blue!50,very thick,fill=blue!10}]
			\node[state]         (q1)	{{$q_{i}$}}; 
			\node[state]         (q2) [below=of q1] {{$q_{i+1}$}};
			\path[->] 
			(q1) edge [right] node  {{$x_n:=x_n+1$}} (q2) ;
		\end{tikzpicture}
		\qquad
		\begin{tikzpicture}[shorten >=0pt,node distance=2cm,on grid,>=stealth',every state/.style={inner sep=0pt, minimum size=7mm, draw=blue!50,very thick,fill=blue!10}]
			\node[state]         (q1)	{{$q_{i}$}}; 
			\node[state]         (q2) [below=of q1] {{$q_{i+1}$}};
			\path[->] 
			(q1) edge [right] node  {{$x_n:=x_m$}} (q2) ;
		\end{tikzpicture}
		\caption{BACFGs simulating: $z(n)$ (left), $s(n)$ (center), $t(m,n)$ (right).}
		\label{fig:turing-completeness-instr}
		\centering
		\bigskip
		\begin{tikzpicture}[shorten >=0pt,node distance=2cm,on grid,>=stealth',every state/.style={inner sep=0pt, minimum size=7mm, draw=blue!50,very thick,fill=blue!10}]
			\node[state]         (q1)	{{$q_{i}$}}; 
			\node[state]         (q2) [below right=2.5cm of q1] {{$inc_i$}};
			\node[state]         (q3) [below =of q2] {{$q_{i+1}$}};
			\node[state]         (qq) [below left=2.5cm of q1] {{$q_{p}$}};
			
			\path[->] 
			(q1) edge [in=70, out=200, left] node  {{$x_m=x_n?$}} (qq) 
			(q1) edge [in=110, out=-20] [right] node {{$z:=x_m+1$}} (q2)
			(q1) edge [in=160, out=-70] [left] node {{$z:=x_n+1$}} (q2)
			(q2) edge [in=110, out=-110] [left] node  {{$x_n=z?$}} (q3)
			(q2) edge [in=70, out=-70] [right] node  {{$x_m=z?$}} (q3)
			(q2) edge [loop right] node {{$z := z+1$}}(q2);
		\end{tikzpicture}
		\caption{BACFG simulating a jump instruction $j(m,n,p)$.}
		\label{fig:turing-completeness-jump}
	\end{figure}
	
	We next give a precise definition of a model of
	computation for BACFGs which is able to
	simulate URMs.
	Firstly, let us formalise the operational semantics of URMs.
	Given a URM program $P=(I_1,\dots,I_t)$ consisting of a sequence of $t$ instructions $I_j$, 
	we denote its states
	by vectors $\Vec{x}\in\mathbb{N}^{k_P}$, where $k_P$ is the largest
	index of registers used by $P$ (which is finite).
	A configuration of a URM is a pair
	$\langle \Vec{x}, c \rangle \in \mathbb{N}^{k_P}\times \mathbb{N}$
	representing the state of the (possibly used) registers, and 
	the current instruction $I_c$. Then, the operational semantics is as follows:
	
	\begin{definition}[{\bf Operational semantics $\Rightarrow$ of URMs}]
		\label{de:operational-semantics-URM}
		Given a URM program $P=(I_1,\dots, I_t)$, its operational semantics is given by the transition function $\Rightarrow:(\mathbb{N}^{k_P}\times \mathbb{N})\to (\mathbb{N}^{k_P}\times \mathbb{N})$ defined as follows: for all $\vec{x}\in\mathbb{N}^{k_P}, 1\leq c\leq t$,
		\[
		\langle \vec{x} , c\rangle
		\Rightarrow\!
		\begin{cases}
			\langle (x_1,..., x_{n-1}, 0, x_{n+1},..., x_{k_P}), c+1\rangle & \text{if } I_c=z(n)\\
			\langle (x_1, ..., x_{n-1}, x_n \!+ \!1,x_{n+1}, ..., x_{k_P}), c+1\rangle\!\!\! & \text{if } I_c=s(n)\\
			\langle (x_1, ...,x_{m-1}, x_n, x_{m+1}, ..., x_{k_P}),c+1\rangle & \text{if } I_c=t(m,n)\\
			\langle \Vec{x}, q\rangle & \text{if } I_c=j(m,n,q) \land x_m=x_n\\
			\langle \Vec{x}, c+1\rangle & \text{if } I_c=j(m,n,q) \land x_m\neq x_n\\
		\end{cases}
		\]
		The URM halts when it reaches a configuration 
		$\langle \vec{x} , t+1\rangle$. 
		\qed
	\end{definition}
	
	Getting back to control flow graphs, let us point out 
	that the collecting semantics of BACFGs of Definition~\ref{de:collecting-semantics} 
	can be expressed in terms of Kleene's iterates as follows. 
	
	\begin{definition}[{\bf Kleene's iterates of BACFGs}]
		\label{de:kleene-iterations}
		Let $G=(N, E, s, e)$ be a BACFG with $k_G$ variables. 
		The corresponding initial state 
		$\bot_{\Vec{x}}^s: N\to\wp(\mathbb{N}^{k_G})$, with $\Vec{x}\in \mathbb{N}^{k_G}$,
		and transformer $F_G:(N\to\wp(\mathbb{N}^{k_G}))\to (N\to\wp(\mathbb{N}^{k_G}))$ are
		defined as follows:
		for all $v\in N$ and $\mathcal{X}\in N\to\wp(\mathbb{N}^{k_G})$,
		\[
		\begin{split}
			\bot_{\Vec{x}}^s[v] &\triangleq
			\begin{cases}
				\{\Vec{x}\} & \text{ if } v=s\\
				\varnothing & \text{ otherwise }
			\end{cases}
			\\
			F_G(\mathcal{X})[v]
			& \triangleq
			\underset{(u,c,v)\in E}{\bigcup}
			f_ c (\mathcal{X}[u]) \cup \mathcal{X}[v]
		\end{split}
		\]
		The sequence of \emph{Kleene's iterates of $G$} starting from $\bot_{\Vec{x}}^s$ is the infinite (pointwise) ascending chain $\{F_G^i(\bot_{\Vec{x}}^s)\}_{i\in \mathbb{N}} \subseteq N\to\wp(\mathbb{N}^{k_G})$, where the powers of the function $F_G$ are inductively defined in the usual way: $F_G^0 (\mathcal{X})\eqd 
		\mathcal{X}$ and $F_G^{i+1}(\mathcal{X}) \eqd F_G(F_G^i(\mathcal{X}))$. 
		\qed
	\end{definition}
	
	Observe that the collecting semantics of
	Definition~\ref{de:collecting-semantics} coincides with the least
	fixed point of $F_G$ above $\bot_{\Vec{x}}^s$ w.r.t.\ the
	pointwise inclusion order of the complete lattice
	$N\to\wp(\mathbb{N}^{k_G})$ obtained by lifting
	$\tuple{\wp(\mathbb{N}^{k_G}),\subseteq}$.  Moreover, since $F_G$ is a
	Scott-continuous function (even more, $F_G$ preserves arbitrary least upper
	bounds), by Kleene's fixpoint theorem, it turns out that
	\begin{equation*}
		\cup _{i\in\mathbb{N}} F_G^{i}(\bot_{\Vec{x}}^s)[v] = \llbracket
		G\rrbracket_{\Vec{x}}[v].
	\end{equation*}
	Our key insight is that the states of our
	abstract computational model
	%, as formalised in the following lemma, 
	can be represented as 
	``differences'' between consecutive Kleene's iterates of $F_G$.
	
	\begin{definition}[{\bf Operational semantics $\Delta$ of BACFGs}]
		\label{de:delta-function}
		Given a BACFG $G=(N, E, s, e)$, its operational semantics is given by the function $\Delta_G:(N\to\wp(\mathbb{N}^{k_G}))\to (N\to\wp(\mathbb{N}^{k_G}))$ defined as follows: for all $\mathcal{X}:N\rightarrow\wp(\mathbb{N}^{k_G})$ and $v\in N$, 
		
		\medskip
		\(\qquad \Delta_G(\mathcal{X})[v]\eqd \underset{(u,c,v)\in E}{\bigcup}
		f_ c (\mathcal{X}[u]).\)
		\qed
	\end{definition}
	
	Thus, $\Delta_G(\mathcal{X})[v]$ is the standard ``meet-over-paths'' of dataflow analysis, namely, 
	the join of the transfer functions $f_c(X)$ 
	over all the edges $(u,c,v)$ of $G$.  
	\smallskip
	
	\begin{lemma}
		\label{le:F-as-delta-sum}
		Let $G=(N,E,s,e)$ be a BACFG. For all $n\in\mathbb{N}$, $\mathcal{X}:N\rightarrow\wp(\mathbb{N}^{k_G})$, $v\in N$, we have that
		$F_G^n(\mathcal{X})[v]=\cup_{0\leq i\leq n} \Delta_G^{i}(\mathcal{X})[v]$.
	\end{lemma}
	\begin{proof}
		We proceed by induction on $n\in \mathbb{N}$.
		\begin{itemize}
			\item $n=0$: $F_G^0(\mathcal{X})[v]=\mathcal{X}[v]=\Delta_G^{0}(\mathcal{X})[v]=\cup_{0\leq i\leq 0} \Delta_G^{i}(\mathcal{X})[v]$;
			\item $n>0$:
			\begin{align*}
				F_G^{n}(\mathcal{X})[v]
				&=
				F_G(F_G^{n-1}(\mathcal{X}))[v]\\
				&=
				\underset{(u,c,v)\in E}{\textstyle\bigcup}
				f_ c (F_G^{n-1}(\mathcal{X})[u]) \cup F_G^{n-1}(\mathcal{X})[v]\tag*{[by ind.\ hyp.]}\\
				&=
				\Delta_G(\cup_{0\leq i\leq n-1} \Delta_G^{i}(\mathcal{X}))[v]\cup (\cup_{0\leq i\leq n-1} \Delta_G^{i}(\mathcal{X})[v])\tag*{[by additivity of $\Delta_G$]}\\
				&=
				\cup_{1\leq i\leq n} \Delta_G^{i}(\mathcal{X})[v]\cup (\cup_{0\leq i\leq n-1} \Delta_G^{i}(\mathcal{X})[v])\\
				&=\cup_{0\leq i\leq n} \Delta_G^{i}(\mathcal{X})[v]
			\end{align*}
		\end{itemize}
		This closes the proof.
	\end{proof}
	
	In the following, we describe an effective procedure $\tau$ to translate a URM program $P$ into a BACFG which simulates $P$. 
	
	\begin{definition}[{\bf Transformer $\tau$}]\label{def:tau}
		Given a URM $P=(I_1,\dots,I_t)$,
		the procedure $\tau(P)$ starts from $N_0=\{q_1, \dots, q_t, q_{t+1} \}$ and $E_0=\varnothing$ as, resp., sets of nodes and edges. Then, for all the instructions $I_i$ of $P$:
		
		\begin{enumerate}[{\rm (i)}]
			\item If $I_i\in \{z(n), s(n), t(m,n)\mid n,m\in\mathbb{N}\}$ then $\tau(P)$ adds an edge between the nodes $q_{i}$ and $q_{i+1}$ as depicted by the diagrams in Fig.~\ref{fig:turing-completeness-instr}. For instance, if $I_i=z(n)$ the edge $(q_i,x_n:=0,q_{i+1})$ is added to the set $E$; the other cases are analogous.
			\item If $I_i=j(m,n,q)$, for some $m,n,q$, then $\tau(P)$ adds a new node $inc_i$ and the edges depicted by the diagram in Fig.~\ref{fig:turing-completeness-jump}. We shall use the variable $z$ as a syntactic shorthand for $x_{k_P+1}$, which  is a fresh variable not used in $P$.
		\end{enumerate}

		Let $N$ and $E$ denote the final sets of, resp., nodes and edges obtained by applying the above two steps 
		(i)--(ii) for all the instructions of $P$. 
		Then, $\tau(P)$ returns a set of BACFGs $\{(N,E,q_s,q_e)\mid q_s,q_e\in N_0\}$, where 
		start and end nodes freely range in $N_0$ and each BACFG has 
		$k_G\in \{k_P,k_P+1\}$ variables.
		%\todo{P: explain that we want to be able to fix input and output node freely, that's why $q_s,q_e$ are generic? L: in realtà il motivo è legato ad alcune dimostrazioni induttive: con $q_1$ e $q_{t+1}$ l'ipotesi induttiva è troppo debole}
		%
		%In what follows, 
		Without loss of generality, we assume $k_G \triangleq k_P+1$: In fact, if the program $P$ 
		contains no jump and the extra-variable $z$ is actually not used, then we can add a useless edge involving the extra-variable $z$.
		\qed
	\end{definition}
	
	In the rest of this section, we prove that the BACFG $G=(N, E, q_1, q_{t+1})\in \tau(P)$ simulates the original URM program $P$.
	To prove our claim, we define an equivalence relation between sets of states of a BACFG in $\tau(P)$. Intuitively, two sets $\mathcal{X}$ and $\mathcal{X}'$ are deemed equivalent if, for each node, $\mathcal{X}$ and $\mathcal{X}'$ induce the same invariant on the first $k_P$ variables, except for the states $inc_i$ whose variable $z$ is already greater than the variables occurring in the outgoing guards.
	
	\begin{definition}[{\bf Equivalence $\approx$}]
		\label{de:almost-equivalent}
		Let $P=(I_1,\dots,I_t)$ be a URM program and $G=(N,E,q_s,q_e)\in \tau(P)$. Then, given $\mathcal{X},\mathcal{X}':N\to\wp(\mathbb{N}^{k_G})$, the relation 
		$\mathcal{X} \approx \mathcal{X}'$ is defined as follows: 
		
		\begin{enumerate}[(1)]
			\item $\forall i\in [1,t+1].\, \mathcal{X}[q_i]\!\restriction_{k_P}=\mathcal{X'}[q_i]\!\restriction_{k_P}$;
			\medskip
			\item $\forall i\in [1,t], \forall m\in [1,k_P], \forall (inc_i,x_m=z?,q_{i+1})\in E$.\,
			\(
			\{\Vec{x}\in \mathcal{X}[inc_i]\mid z\leq x_m\}=\{\Vec{x}\in\mathcal{X}'[inc_i]\mid z\leq x_m\}
			\).
			\medskip \qed
		\end{enumerate}
	\end{definition}
	
	Let us point out that condition (2) is motivated by the observation that for nodes of type $inc_i$, the states containing values of $x_m$ below $z$ do not matter.
	%only whenever the value of $z$ is above $x_m$, differences the value affect the semantics.
	Clearly, observe that $\approx$ is an equivalence relation. Moreover, it turns out that the operational semantic function $\Delta_G$ of Definition~\ref{de:delta-function} preserves this equivalence~$\approx$ .
	\begin{lemma}
		\label{le:delta-preserve-approx}
		Let $P=(I_1,\dots,I_t)$ be a URM program and $G=(N,E,q_s,q_e)\in \tau(P)$. Then, for all  $\mathcal{X},\mathcal{X}':N\to\wp(\mathbb{N}^{k_G})$,
		\(
		\mathcal{X}\approx \mathcal{X'} \Rightarrow 
		\Delta_G(\mathcal{X})\approx\Delta_G(\mathcal{X}')
		\).
	\end{lemma}
	\begin{proof} Assume that $\mathcal{X}\approx \mathcal{X'}$. 
		For all $i\in [1,t+1]$ we have:
		\begin{align*}
			& \Delta_G(\mathcal{X})[q_i]
			\restriction_{k_P} \\                       
			& = 		
			\textstyle\underset{(u,c,q_i)\in E}{\bigcup} f_c (\mathcal{X}[u])
			\restriction_{k_P}\\
			& =  
			\textstyle\underset{(q_u,c,q_i)\in E}{\bigcup} f_c (\mathcal{X}[q_u])\restriction_{k_P}
			\cup
			\textstyle\underset{(inc_{i-1},x_m=z?,q_i)\in E}{\bigcup} f_{x_m=z?} (\mathcal{X}[inc_{i-1}]) \restriction_{k_P}
			\\
			& =  
			\textstyle\underset{(q_u,c,q_i)\in E}{\bigcup} f_c (\mathcal{X}[q_u])\restriction_{k_P} 
			\cup
			\textstyle\underset{(inc_{i-1},x_m=z?,q_i)\in E}{\bigcup} f_{x_m=z?} (\{\Vec{x}\in\mathcal{X}[inc_{i-1}]\mid z\leq x_m\}) \restriction_{k_P}
			\tag*{[as $\mathcal{X}\approx\mathcal{X}'$]}
			\\
			&=
			\textstyle\underset{(q_u,c,q_i)\in E}{\bigcup} f_c (\mathcal{X}'[q_u]) \restriction_{k_P}
			\cup
			\textstyle\underset{(inc_{i-1},x_m=z?,q_i)\in E}{\bigcup} f_{x_m=z?} (\{\Vec{x}\in\mathcal{X}'[inc_{i-1}]\mid z\leq x_m\})\restriction_{k_P}
			\\             
			&=                        
			\textstyle\underset{(q_u,c,q_i)\in E}{\bigcup} f_c (\mathcal{X}'[q_u]) \restriction_{k_P}
			\cup
			\textstyle\underset{(inc_{i-1},x_m=z?,q_i)\in E}{\bigcup} f_{x_m=z?} (\mathcal{X}'[inc_{i-1}]) \restriction_{k_P}
			\\
			& =\Delta_G(\mathcal{X}')[q_i] \restriction_{k_P}.
		\end{align*}
		Moreover, for all $i\in [1,t]$, $m\in [1,k_P]$ such that $(inc_i,x_m=z?,q_{i+1})\in E$:
		\begin{align*}
			&\{\Vec{x}\in \Delta_G(\mathcal{X})[inc_i]\mid z\leq x_m\}\\
			&=\{\Vec{x}\in \textstyle\underset{(u,c,inc_i)\in E}{\bigcup} f_c (\mathcal{X}[u])\mid z\leq x_m\}\\
			&=\{\Vec{x}\in\textstyle\underset{(q_i,c,inc_i)\in E}{\bigcup}f_c (\mathcal{X}[q_i]) \mid z\leq x_m \}\cup \{\vec{x}\in f_{z:=z+1} (\mathcal{X}[inc_{i}])\mid z\leq x_m\}\\
			&=\{\Vec{x}\in f_{z:=x_n+1} (\mathcal{X}[q_i])\cup f_{z:=x_m+1} (\mathcal{X}[q_i])  \mid z\leq x_m \}\cup\\
			&\hspace*{45ex} \{\vec{x}\in f_{z:=z+1} (\mathcal{X}[inc_{i}])\mid z\leq x_m\}\\
			&=\{\Vec{x}\in f_{z:=x_n+1} (\mathcal{X}[q_i])\mid z\leq x_m \}\cup \{\vec{x}\in f_{z:=z+1} (\mathcal{X}[inc_{i}])\mid z\leq x_m\},
		\end{align*}
		for some $n\neq m$. Since $\mathcal{X}[q_i]\restriction_{k_P}=\mathcal{X}'[q_i]\restriction_{k_P}$ it follows that $ f_{z:=x_n+1} (\mathcal{X}[q_i])= f_{z:=x_n+1} (\mathcal{X}'[q_i])$.
		Also note that:
		\begin{align*}
			&\{\vec{x}\in f_{z:=z+1} (\mathcal{X}[inc_{i}])\mid z\leq x_m\}\\
			&=
			\{\vec{x}\in f_{z:=z+1} (\{\vec{x}\in\mathcal{X}[inc_{i}]\mid z\leq x_m\})\mid z\leq x_m\}\tag*{[as $\mathcal{X}\approx \mathcal{X}'$]}\\
			&=		\{\vec{x}\in f_{z:=z+1} (\{\vec{x}\in\mathcal{X}'[inc_{i}]\mid z\leq x_m\})\mid z\leq x_m\} \\
			&=	\{\vec{x}\in f_{z:=z+1} (\mathcal{X}'[inc_{i}])\mid z\leq x_m\}.
		\end{align*}
		Hence,
		\begin{align*}
			&\{\Vec{x}\in \Delta_G(\mathcal{X})[inc_i]\mid z\leq x_m\}\\
			&=\{\Vec{x}\in f_{z:=x_n+1} (\mathcal{X}[q_i])\mid z\leq x_m \}\cup \{\vec{x}\in f_{z:=z+1} (\mathcal{X}[inc_{i}])\mid z\leq x_m\}\\
			&=\{\Vec{x}\in f_{z:=x_n+1} (\mathcal{X}'[q_i])\mid z\leq x_m \}\cup \{\vec{x}\in f_{z:=z+1} (\mathcal{X}'[inc_{i}])\mid z\leq x_m\}\\
			&=\{\Vec{x}\in\Delta_G(\mathcal{X}')[inc_i]:  z\leq x_m\}.
		\end{align*}
		This therefore shows that $\Delta_G(\mathcal{X})\approx \Delta_G(\mathcal{X}')$.
	\end{proof}
	
	Let us now show that each transition of a URM program can be simulated by a finitely many applications, say $k$, of the function $\Delta$. Moreover, whenever $\Delta$ is applied less than $k$ times, we obtain the empty set of states for all the nodes. Let us define the following concatenation operation for sequences: $(a_1,\dots,a_k):a\triangleq (a_1,\dots,a_k,a)$. Concatenation will be used to deal with the fact that our transformed BACFG has  an additional variable w.r.t.\ 
	the original URM program.
	
	\begin{lemma}
		\label{le:delta-1-simulation}
		Let $P=(I_1,\dots,I_t)$ be a URM program. For all BACFGs $G=(N,E,q_s,q_e)\in \tau(P)$, $\Vec{x}, \Vec{x}'\in \mathbb{N}^{k_P}$,  $s'\in \mathbb{N}$, if $\langle \Vec{x},s\rangle \Rightarrow \langle \Vec{x}', s '\rangle$ then there exists $k\in\mathbb{N}$ such that:
		\begin{enumerate}[(1)]
			\item $\Delta_G^k(\bot_{\Vec{x}:0}^{q_s})\approx \bot_{\Vec{x}':0}^{q_{s'}}$;
			\item $\forall i\in [1,k-1], \forall j\in [1,t+1].\:  \Delta_G^i(\bot_{\Vec{x}:0}^{q_s})[q_j]=\varnothing$.
		\end{enumerate}
	\end{lemma}
	\begin{proof}
		Assume that $\langle \Vec{x},s\rangle \Rightarrow \langle \Vec{x}', s '\rangle$. 
		We distinguish three cases.
		
		\medskip
		\noindent
		(i) Let $I_{s}\in \{z(n), s(n), t(m,n) \mid n,m\in\mathbb{N}\}$. Consider the case $I_s=z(n)$ for some $n$ (the remaining cases are analogous), so that $s'=s+1$. For $k=1$ we have that:
		\begin{align*}
			\Delta_G(\bot_{\Vec{x}:0}^{q_s})&=
			\lambda v.\ \textstyle\underset{(u,c,v)\in E}{\bigcup} f_c (\bot_{\Vec{x}:0}^{q_s}[u])\\
			&=\lambda v.
			\begin{cases}
				f_{x_n:=0} (\{\Vec{x}:0\}) & \text{ if } v=q_{s+1}\\
				\varnothing & \text{ otherwise }
			\end{cases}\tag*{[by def.\ of $G$]}\\
			&=\lambda v.
			\begin{cases}
				\Vec{x}':0 & \text{ if } v=q_{s+1}\\
				\varnothing & \text{ otherwise }
			\end{cases}\\
			&= \bot_{\Vec{x}':0}^{q_{s'}} \tag*{[as $s'=s+1$]}
		\end{align*}
		Thus, $\Delta_G(\bot_{\Vec{x}:0}^{q_s})\approx \bot_{\Vec{x}':0}^{q_{s'}}$, i.e., property 
		(1) holds with $k=1$. 
		Property (2) trivially holds since for $k=1$, $[1,k-1]$ is the empty set. 
		
		\medskip
		\noindent		
		(ii)		 Let $I_{s}=j(m,n,p)$ and assume that $x_m=x_n$ holds, so that the next instruction to execute is $I_q$, i.e., $s'=p$. For $k=1$ we have that:
		\begin{align*}
			\Delta_G(\bot_{\Vec{x}:0}^{q_s})&=
			\lambda v.\ \textstyle\underset{(u,c,v)\in E}{\bigcup} f_c (\bot_{\Vec{x}:0}^{q_s}[u])\\
			&=\lambda v.
			\begin{cases}
				f_{x_m=x_n?} (\{\Vec{x}:0\}) & \text{ if } v=q_{p}\\
				f_{z:=x_m+1} (\{\Vec{x}:0\})\cup f_{z:=x_n+1} (\{\Vec{x}:0\}) & \text{ if } v=inc_{s}\\
				\varnothing & \text{ otherwise }
			\end{cases}\tag*{[by def.\ of $G$]}\\
			&=\lambda v.
			\begin{cases}
				\{\Vec{x}:0\} & \text{ if } v=q_{p}\\
				\{\Vec{x}:x_m+1\}& \text{ if } v=inc_{s}\\
				\varnothing & \text{ otherwise }
			\end{cases} \tag*{[as $x_m=x_n$]}
		\end{align*}
		Since $s'=p$ and $\vec{x}=\vec{x}'$, we have that:
		\begin{itemize}
			\item for all $i\in[1,t+1], \Delta_G(\bot_{\Vec{x}:0}^{q_s})[q_i]=\bot_{\Vec{x}:0}^{q_{p}}[q_i]=\bot_{\Vec{x}':0}^{q_{s'}}[q_i]$;
			\item for all $i\in[1,t+1]$ and $m\in[1,k_P]$ such that $(inc_i,x_m=z?,q_{i+1})\in E$:
			\[\{\vec{x}\in \Delta_G(\bot_{\Vec{x}:0}^{q_s})[inc_i] \mid z\leq x_m \}=\varnothing=\{\vec{x}\in 
			\bot_{\Vec{x}':0}^{q_{s'}}[inc_i] \mid z\leq x_m \}.\]
		\end{itemize}
		Thus, $	\Delta_G(\bot_{\Vec{x}:0}^{q_s}) \approx \bot_{\Vec{x}':0}^{q_{s'}}$ holds, 
		i.e., property~(1) holds with $k=1$. Moreover,
		once again property (2) trivially holds because $[1,k-1]$ is empty. 
		
		\medskip
		\noindent
		(iii) The last possible case is
		$I_{s}=j(m,n,q)$ with $x_m\neq x_n$, so that the next instruction to execute is $I_{s+1}$, i.e., 
		$s'=s+1$. We first prove, by induction, that for all $i\geq 1$:
		
		\begin{multline}\label{eq:jump-delta}
			i\leq |x_m-x_n| \Rightarrow \\
			\Delta_G^i(\bot_{\Vec{x}:0}^{q_s})= 
			\lambda v.
			\begin{cases}
				f_{z:=x_n+i} (\{\Vec{x}:0\})\cup f_{z:=x_m+i} (\{\Vec{x}:0\}) & \text{if } v=inc_{s}\\
				\varnothing & \text{otherwise }
			\end{cases}\tag{$*$}
		\end{multline}
		
		\noindent
		For the base case $i=1$, we have that:
		\begin{align*}
			&\Delta_G(\bot_{\Vec{x}:0}^{q_s})\\
			&=\lambda v.\ \textstyle\underset{(u,c,v)\in E}{\bigcup} f_c (\bot_{\Vec{x}:0}^{q_s}[u])\\
			&=
			\lambda v.
			\begin{cases}
				f_{z:=x_n+1} (\{\Vec{x}:0\})\cup f_{z:=x_m+1} (\{\Vec{x}:0\}) & \text{ if } v=inc_{s}\\
				\varnothing & \text{ otherwise }
			\end{cases}
			\tag*{[by def.\ of $G$]}
		\end{align*}
		For the inductive case $i>1$, assume that $i\leq |x_m-x_n|$ (if $i>|x_m-x_n|$ the implication \eqref{eq:jump-delta} trivially holds). We have that:
		\begin{align*}
			&\Delta_G^i(\bot_{\Vec{x}:0}^{q_s})\\[5pt]
			&=\Delta_G (\Delta_G^{i-1}(\bot_{\Vec{x}:0}^{q_s}))\\[5pt]
			&=\Delta_G \left( \lambda v.
			\begin{cases}
				f_{z:=x_n+i-1} (\{\Vec{x}:0\})\cup f_{z:=x_m+i-1} (\{\Vec{x}:0\}) & \text{ if } v=inc_{s}\\
				\varnothing & \text{ otherwise }
			\end{cases} \right) \tag*{[by ind.\ hyp.\ for $i-1\leq |x_m-x_n|$]}
			\\[5pt]
			&=
			\lambda v.
			\begin{cases}
				f_{z:=z+1} \big(f_{z:=x_n+i-1} (\{\Vec{x}:0\})\cup f_{z:=x_m+i-1} (\{\Vec{x}:0\})\big) & \text{ if } v=inc_{s}\\
				\varnothing & \text{ otherwise }
			\end{cases} \\
			\tag*{[as $(inc_s,z:=z+1,inc_s)$ is an edge of $G$ and}\\
			\tag*{$x_m\neq x_n+i-1$ and $x_n \neq x_m+i-1$ since $i-1<|x_m-x_n|$]}
			\\[5pt]
			&=
			\lambda v.
			\begin{cases}
				f_{z:=x_n+i} (\{\Vec{x}:0\})\cup f_{z:=x_m+i} (\{\Vec{x}:0\}) & \text{ if } v=inc_{s}\\
				\varnothing & \text{ otherwise }
			\end{cases}
		\end{align*}
		We have therefore shown the implication \eqref{eq:jump-delta}. Now, note that for $k=|x_m-x_n|+1$ we have that:
		\begin{align*}
			&\Delta_G^{|x_m-x_n|+1}(\bot_{\Vec{x}:0}^{q_s})\\
			&=\Delta_G(\Delta_G^{|x_m-x_n|}(\bot_{\Vec{x}:0}^{q_s}))\\
			&=\Delta_G
			\left(
			\lambda v.
			\begin{cases}
				f_{z:=x_n+|x_m-x_n|} (\{\Vec{x}:0\})\cup f_{z:=x_m+|x_m-x_n|} (\{\Vec{x}:0\}) & \text{ if } v=inc_{s}\\
				\varnothing & \text{ otherwise }
			\end{cases}\right)  \tag*{[by~\eqref{eq:jump-delta}]}\\
			&=\lambda v.
			\begin{cases}
				f_{z:=x_n+|x_m-x_n|+1} (\{\Vec{x}:0\})\cup f_{z:=x_m+|x_m-x_n|+1} (\{\Vec{x}:0\}) & \text{ if } v=inc_{s}\\
				\{\Vec{x}:\max(x_m,x_n)\} & \text{ if } v=q_{s+1}\\
				\varnothing & \text{ otherwise }
			\end{cases}
			\tag*{[because $\max(x_m,x_n)=\min(x_m,x_n)+|x_m-x_n|$]}
		\end{align*}
		Since $s'=s+1$ and $\vec{x}=\vec{x}'$, we have that:
		\begin{itemize}
			\item for all $i\in [1,t+1]$, $\Delta_G^{|x_m-x_n|+1}(\bot_{\Vec{x}:0}^{q_s})[q_i]\!\restriction_{k_P}=\bot_{\Vec{x}:0}^{q_{s+1}}[q_i]\!\restriction_{k_P}=\bot_{\Vec{x}':0}^{q_{s'}}[q_i]\!\restriction_{k_P}$;
			\item for all $i\in [1,t+1]$ and $m\in [1, k_P]$ 
			such that $(inc_i,x_m=z?,q_{i+1})\in E$:
			\[\{\vec{x}\in \Delta_G^{|x_m-x_n|+1}(\bot_{\Vec{x}:0}^{q_s})[inc_i] \mid z\leq x_m \}=\varnothing=\{\vec{x}\in 
			\bot_{\Vec{x}':0}^{q_{s'}}[inc_i] : z\leq x_m \}.\]
		\end{itemize}
		Therefore, $	\Delta_G(\bot_{\Vec{x}:0}^{q_s}) \approx \bot_{\Vec{x}':0}^{q_{s'}}$ holds. 
		Furthermore, for all $i\in [1,|x_m-x_n|]$, by applying the implication~\eqref{eq:jump-delta} we obtain:
		\[\Delta_G^i(\bot_{\Vec{x}:0}^{q_s})=
		\begin{cases}
			f_{z:=x_n+i} (\{\Vec{x}:0\})\cup f_{z:=x_m+i} (\{\Vec{x}:0\}) & \text{ if } v=inc_{s}\\
			\varnothing & \text{ otherwise.}
		\end{cases}
		\]
		Thus, for all $j\in [1,t+1]$, $\Delta_G^i(\bot_{\Vec{x}:0}^{q_s})[q_j]=\varnothing$ holds and this concludes the proof.
	\end{proof}
	
	Let us now generalise Lemma \ref{le:delta-1-simulation} to any number of execution steps $\Rightarrow^n$ performed by a URM program. In particular, we show that if the URM halts then our abstract model will reach, after finitely many steps, a state that stores the URM output in its end node. Likewise, whenever the URM diverges, the state of the end node will be empty.
	
	\begin{proposition}
		\label{prop:delta-n-simulation}
		Let $P=(I_1,\dots,I_t)$ be a URM program. Then, for all $G=(N,E,q_s,q_e)\in \tau(P)$, $\Vec{x},\Vec{x}'\in \mathbb{N}^{k_P}$, $n\in\mathbb{N}$, 
		if $\langle \Vec{x}, s \rangle \Rightarrow^n \langle \Vec{x}', t+1 \rangle$ then
		there exists $n'\in \mathbb{N}$ such that: 
		\begin{enumerate}[(1)]
			\item $\Delta_G^{n'}(\bot_{\Vec{x}:0}^{q_s}) \approx \bot_{\Vec{x}':0}^{q_{t+1}}$;
			\item $\forall i\in [0,n'-1].\: \Delta_G^{i}(\bot_{\Vec{x}:0}^{q_s})[q_{t+1}]= \varnothing$.
		\end{enumerate}
	\end{proposition}
	\begin{proof}
		We proceed by induction on $n\in \mathbb{N}$.
		\begin{itemize}
			\item Base case $n=0$, so that $\langle \Vec{x},s\rangle=\langle \Vec{x}', t+1 \rangle$. Therefore, for $n'=0$ the property (1) holds because:
			\begin{align*}
				\Delta_G^{n'}(\bot_{\Vec{x}:0}^{q_s})
				&=\Delta_G^{0}(\bot_{\Vec{x}':0}^{q_{t+1}}) \tag*{[as $n'=0,t+1=s$, $\Vec{x}'=\Vec{x}$]}\\
				&=\bot_{\Vec{x}':0}^{q_{t+1}} \tag*{[as $\Delta_G^0=\lambda x .x$]}
			\end{align*}
			Moreover, the property (2) trivially holds becase $[0,n'-1]$ is empty. 
			\item Inductive case $n>0$, so that $\langle \Vec{x}, s\rangle \Rightarrow \langle \Vec{x}'', s'' \rangle \Rightarrow^{n-1} \langle \Vec{x}', t+1 \rangle$. We have that:
			\begin{itemize}
				\item by Lemma \ref{le:delta-1-simulation}, and observing that $s\neq t+1$, we know that there exists $m\in \mathbb{N}$ such that: (1)~$\Delta_G^{m}(\bot_{\Vec{x}:0}^{q_s})\approx \bot_{\Vec{x}'':0}^{q_{s''}}$; (2)~$\forall i\in[0,m-1].\ \Delta_G^{i}(\bot_{\Vec{x}:0}^{q_{s}})[q_{t+1}]=\varnothing$.
				
				\item by inductive hypothesis there exists $n''\in \mathbb{N}$ such that: (i)~$\Delta_G^{n''}(\bot_{\Vec{x}'':0}^{q_{s''}})\approx \bot_{\Vec{x}':0}^{q_{t+1}}$; (ii)~$\forall i\in[0,n''-1].\ \Delta_G^{i}(\bot_{\Vec{x}'':0}^{q_{s''}})[q_{t+1}]= \varnothing$.
			\end{itemize}
			Therefore, it turns out that:
			\begin{align*}
				\Delta_G^{n''+m}(\bot_{\Vec{x}:0}^{q_s})
				&=\Delta_G^{n''}(\Delta_G^m(\bot_{\Vec{x}:0}^{q_s})) \\
				&\approx\Delta_G^{n''}(\bot_{\Vec{x}'':0}^{q_{s''}}) \tag*{[as $\Delta_G^{m}(\bot_{\Vec{x}:0}^{q_s})\approx \bot_{\Vec{x}'':0}^{q_{s''}}$, by Lemma \ref{le:delta-preserve-approx}]}\\
				&\approx \bot_{\Vec{x}':0}^{q_{t+1}} \tag*{[by ind.\ hyp.]}
			\end{align*}
			thus showing (1) for $n''+m$. 
			Moreover, for all $i\in [0,n''-1]$:
			\begin{align*}
				\Delta_G^{i+m}(\bot_{\Vec{x}:0}^{q_s})
				&=\Delta_G^{i}(\Delta_G^{m}(\bot_{\Vec{x}:0}^{q_s}))\\
				&\approx \Delta_G^i(\bot_{\Vec{x}'':0}^{q_{s''}}). \tag*{[as $\Delta_G^{m}(\bot_{\Vec{x}:0}^{q_s})\approx \bot_{\Vec{x}'':0}^{q_{s''}}$, by Lemma \ref{le:delta-preserve-approx}]}
			\end{align*}
			Recall that, by inductive hypothesis, $\Delta_G^i(\bot_{\Vec{x}'':0}^{q_{s''}})[q_{t+1}]=\varnothing$, so that we obtain that for all $i\in[m,n''+m-1]$, $\Delta_G^{i}(\bot_{\Vec{x}:0}^{q_s})[q_{t+1}]=\varnothing$ holds. Since, by Lemma \ref{le:delta-1-simulation}, we have that for all $i\in[0,m-1]$, $\Delta_G^{i}(\bot_{\Vec{x}:0}^{q_{s}})[q_{t+1}]=\varnothing$ holds, we conclude that for all $i\in[0,n''+m-1]$, $\Delta_G^{i}(\bot_{\Vec{x}:0}^{q_{s}})[q_{t+1}]=\varnothing$, thus showing (2) for  $n''+m$. 
			\qedhere
		\end{itemize}
	\end{proof}
	
	\begin{proposition}
		\label{prop:delta-n-simulation-reverse}
		Let $P=(I_1,\dots,I_t)$ be a URM program. Then, for all $G=(N,E,q_s,q_e)\in \tau(P)$, $\Vec{x}\in \mathbb{N}^{k_P}$, $n\in\mathbb{N}$:
		\[
		\text{if }\Delta_G^{n}(\bot_{\Vec{x}:0}^{q_{s}})[q_{t+1}] \neq \varnothing 
		\text{ then } \exists \Vec{x}'\in \mathbb{N}^{k_P}, \exists n'\in \mathbb{N}.\: \langle \Vec{x}, s \rangle \Rightarrow^{n'} \langle \Vec{x}', t+1 \rangle .
		\]
	\end{proposition}
	\begin{proof} We proceed by induction on $n\in \mathbb{N}$:
		\begin{itemize}
			\item $n=0$: by hypothesis, $\Delta_G^{0}(\bot_{\Vec{x}:0}^{q_s})[q_{t+1}]=\bot_{\Vec{x}:0}^{q_s}[q_{t+1}]\neq \varnothing$, so that $s=t+1$ and, in turn,  $\langle \Vec{x}, s \rangle \Rightarrow^0 \langle \Vec{x}, t+1 \rangle$.
			\item $n>0$: by hypothesis, we have that $\Delta_G^{n}(\bot_{\Vec{x}:0}^{q_s})[q_{t+1}]\neq \varnothing$. We consider $s\neq t+1$, otherwise, one can trivially pick $n'=0$. By construction, there exist $\Vec{x}'', s''$ such that $\langle \Vec{x},s\rangle \Rightarrow \langle \Vec{x}'', s''\rangle$, and, by Lemma \ref{le:delta-1-simulation}, there exists $m$ such that $\Delta_G^m(\bot_{\Vec{x}:0}^{q_s})\approx \bot_{\Vec{x}'':0}^{q_{s''}}$. Note that $n\geq m$ holds, since for all $i\in[1,m-1]$,  $\Delta_G^i(\bot_{\Vec{x}:0}^{q_s})[q_{t+1}]=\varnothing$ holds. By Lemma \ref{le:delta-preserve-approx}, it follows that $\Delta_G^{n-m}(\Delta_G^m(\bot_{\Vec{x}:0}^{q_s})) \approx \Delta_G^{n-m}(\bot_{\Vec{x}'':0}^{q_{s''}})$,  By hypothesis and Definition \ref{de:almost-equivalent}, we have that $\Delta_G^{n-m}(\Delta_G^m(\bot_{\Vec{x}:0}^{q_s}))[q_{t+1}]=\Delta_G^{n-m}(\bot_{\Vec{x}'':0}^{q_{s''}})[q_{t+1}]\neq \varnothing$. We conclude by applying the inductive hypothesis, that entails the existence of $m'$ such that $\langle \Vec{x},s\rangle \Rightarrow \langle \Vec{x}'', s'' \rangle \Rightarrow^{m'} \langle \Vec{x}', s' \rangle$. \qedhere
		\end{itemize}
	\end{proof}

	The next two results show that for a given URM program $P=(I_1,\dots,I_t)$, the BACFG $G=(N,E,q_1,q_t)\in\tau(P)$ simulates the operational semantics of $P$ starting from its first instruction $I_1$.
	\begin{proposition}
		\label{prop:urm->cfg-halt}
		Let $P=(I_1,\dots,I_t)$ be a given URM program and $G=(N, E, q_1, q_{t+1})\in \tau(P)$. Then, for all $\Vec{x},\Vec{x}'\in \mathbb{N}^{k_P}$ and $n\in\mathbb{N}$:
		\[
		\text{if }\langle \Vec{x}, 1 \rangle \Rightarrow^n \langle \Vec{x}', t+1 \rangle
		\text{ then }
		\llbracket G \rrbracket_{\Vec{x}:0}[q_{t+1}]\restriction_{k_P} = \{\Vec{x}'\}.
		\]
	\end{proposition}
	\begin{proof}
		By Proposition \ref{prop:delta-n-simulation} there exists $n'$ such that $\Delta_G^{n'}(\bot_{\Vec{x}:0}^{q_1}) \approx \bot_{\Vec{x}':0}^{q_{t+1}}$ and for all $i\in[0,n'-1]$, $\Delta_G^{i}(\bot_{\Vec{x}:0}^{q_1})[q_{t+1}]= \varnothing$. Let us prove, by induction on $i$, that for all $i>n'$,
		$\Delta_G^{i}(\bot_{\Vec{x}:0}^{q_1})\approx \lambda v.\varnothing$. 
		\begin{itemize}
			\item $i=n'+1$:
			\begin{align*}
				\Delta_G^{n'+1}(\bot_{\Vec{x}:0}^{q_1})
				&=\Delta_G(\Delta_G^{n'}(\bot_{\Vec{x}:0}^{q_1}))\\
				&\approx \Delta_G(\bot_{\Vec{x}':0}^{t+1}) \tag*{[as $\Delta_G^{n'}(\bot_{\Vec{x}:0}^{q_1}) \approx \bot_{\Vec{x}':0}^{t+1}$, by Lemma \ref{le:delta-preserve-approx}]}\\
				&=\lambda v.\varnothing. \tag*{[by def.\ of $G$]}
			\end{align*}
			\item $i>n'+1$:
			\begin{align*}
				\Delta_G^{i}(\bot_{\Vec{x}:0}^{q_1})
				&=\Delta_G(\Delta_G^{i-1}(\bot_{\Vec{x}:0}^{q_1}))\\
				&\approx \Delta_G(\lambda v.\varnothing). \tag*{[by ind.\ hyp.\ and Lemma \ref{le:delta-preserve-approx}]}\\
				&= \lambda v.\varnothing
			\end{align*}
		\end{itemize}
		Thus, for all $i\neq n'$, we have that $\Delta_G^{i}(\bot_{\Vec{x}:0}^{q_1})[q_{t+1}]= \varnothing$. Therefore:
		\begin{align*}
			\llbracket G \rrbracket_{\Vec{x}:0}[q_{t+1}]\!\restriction_{k_P}
			&=
			\cup _{i\in\mathbb{N}} F^{i}(\bot_{\Vec{x}:0}^{q_1})[q_{t+1}]\!\restriction_{k_P} \tag*{[by Kleene's fixpoint theorem]}\\
			&=
			\cup_{i\in\mathbb{N}} \Delta_G^i(\bot_{\Vec{x}:0}^{q_1})[q_{t+1}]\!\restriction_{k_P} \tag*{[by Lemma \ref{le:F-as-delta-sum}]}
			\\
			&=\Delta_G^{n'}(\bot_{\Vec{x}:0}^{q_1})[q_{t+1}]\!\restriction_{k_P} \tag*{[as $\forall i\neq n'.\,  \Delta_G^{i}(\bot_{\Vec{x}:0}^{q_1})[q_{t+1}]= \varnothing$]}\\
			&=\{\Vec{x}'\}. \tag*{[as $\Delta_G^{n'}(\bot_{\Vec{x}:0}^{q_1}) \approx \bot_{\Vec{x}':0}^{t+1}$]}
		\end{align*}
		This therefore closes the proof.
	\end{proof}
	
	\begin{proposition}
		\label{prop:urm->cfg-diverge}
		Let $P=(I_1,\dots,I_t)$ be a given URM program and $G=(N, E, q_1, q_{t+1})\in \tau(P)$. Then,
		for all $\Vec{x}\in\mathbb{N}^{k_P}$:
		\[
		\text{if for all }
		\Vec{x}'\in \mathbb{N}^{k_P}, n\in \mathbb{N},\, \langle \Vec{x}, 1 \rangle \not\Rightarrow^n \langle \Vec{x}', t+1 \rangle
		\text{ then }
		\llbracket G \rrbracket_{\Vec{x}:0}[q_{t+1}] = \varnothing.
		\]
	\end{proposition}
	\begin{proof}
		For all $n'\in\mathbb{N}$, by Proposition \ref{prop:delta-n-simulation-reverse}, we have that $\Delta_G^{n'}(\bot_{\Vec{x}:0}^{q_1})[q_{t+1}]= \varnothing$ holds. As a consequence:
		\begin{align*}
			\llbracket G \rrbracket_{\Vec{x}:0}[q_{t+1}]
			&=
			\cup _{i\in\mathbb{N}} F^{i}(\bot_{\Vec{x}:0}^{q_1})[q_{t+1}] \tag*{[by Kleene's fixpoint theorem]}\\
			&=
			\cup_{i\in\mathbb{N}} \Delta_G^i(\bot_{\Vec{x}:0}^{q_1})[q_{t+1}] \tag*{[by Lemma \ref{le:F-as-delta-sum}]}
			\\
			&=\varnothing.  \tag*{\qedhere}
		\end{align*}
	\end{proof}
	
	We are now in position to prove the main result of this section. 
	
	\cfgTuringCompleteness*
	\begin{proof}
		This  follows from Propositions~\ref{prop:urm->cfg-halt} and \ref{prop:urm->cfg-diverge} and Turing completeness of URMs~\cite[Theorem~4.7]{Cut80}.
	\end{proof}

	\subsection{Concrete and Abstract Semantics}
	A key insight is that our concrete semantics is given by URM programs that satisfy the Assumption~\ref{ass-tc} of Turing completeness, while BACFGs provide the abstract semantics. Let us consider two G\"odel numberings for URMs and BACFGs, so that for an index $a\in\mathbb{N}$, 
	$\mathit{RM}_a$ and $G_a$ denote, resp., the $a$-th URM and 
	BACFG programs. 
	The concrete semantics 
	$\langle \phi, =\rangle$ for URMs 
	is defined as follows:
	for any index $a \in \mathbb{N}$, 
	arity $n\in \mathbb{N}$, and input $\vec{x}\in \mathbb{N}^n$,
	\begin{equation}\label{eq:urm-concrete-semantics}
		\phi_a^{(n)}(\vec{x}) \eqd \\
		\begin{cases}
			y\!\! & \text{if } \mathit{RM}_a \text{ on input } \vec{x} \text{ halts with } y \text{ stored on its 1st register}\\
			\uparrow  & \text{otherwise.}
		\end{cases}
	\end{equation}
	
	On the other hand, the abstract semantics of BACFGs is defined as follows.
	\begin{definition}[{\bf State semantics of BACFGs}]
		\label{def:cfg-relational-semantics}
		Let $Q \subseteq \wp({\mathbb{N}^t})$ be a predicate on sets of states with 
		$t\in \mathbb{N}$ variables. The \emph{state semantics} $\langle {Q}, =\rangle$ of BACFGs,  
		for any index $a\in \mathbb{N}$ and 
		arity $n \in \mathbb{N}$,  
		is given by the function ${Q}_a^{(n)} : \mathbb{N}^n \to \{0,1\}$ 
		defined as follows: for all input
		$\vec{x}\in \mathbb{N}^n$, 
		\[
		{{Q}}_a^{(n)}(\vec{x}) \eqd
		\begin{cases}
			1  & \text{ if } \llbracket G_a\rrbracket_{\Vec{x}}[e_a] \neq \varnothing \wedge \mathbin{\llbracket G_a\rrbracket_{\Vec{x}}[e_a]\!\restriction _t} \in Q\\
			0  & \text{ if } \llbracket G_a\rrbracket_{\Vec{x}}[e_a] \neq \varnothing \wedge \mathbin{\llbracket G_a\rrbracket_{\Vec{x}}[e_a]\!\restriction _t} \not\in Q\\
			\uparrow  & \text{ if } \llbracket G_a\rrbracket_{\Vec{x}}[e_a] = \varnothing,
		\end{cases} 
		\]
		where $e_a$ is the end node of the BACFG $G_a$. \qed
	\end{definition}

	Predicates of type $Q \subseteq \wp({\mathbb{N}^t})$ are also known as hyperproperties~\cite{cs10} in program security 
	and 
	the state semantics of Definition~\ref{def:cfg-relational-semantics} models the validity of a given  
	predicate $Q$ at the end node of a BACFG. Note that, from a computability perspective, it is not restrictive to 
	focus on the end node, since this can be arbitrarily chosen in a BACFG. 
	
	\begin{theorem}
		\label{th:state-semantics-properties}
		The state semantics of BACFGs in Definition~\ref{def:cfg-relational-semantics} is ssmn, branching and discharging.
	\end{theorem}
	
	We split the proof of Theorem \ref{th:state-semantics-properties} into
	three separate results that are given below.
	%\todo{P: c'era ``For notational convenience, we use $:$ to denote concatenation of values and vectors as $x:\vec{x}\triangleq (x,x_1,\dots,x_k)$ for all $x\in\mathbb{N}$ and $\vec{x}=(x_1,\dots,x_k)\in \mathbb{N}^*$.''. Mi pare gia' detto prima.}
	In BACFGs, we write the command
	$[x_a,x_{a+i}]:=[x_{b},x_{b+i}]$, for some indices $a$, $b$ and
	$i\geq 0$, to denote a sequence of adjacent edges with commands
	$x_a:=x_{b}$, $x_{a+1}:=x_{b+1}$, $\dots$, $x_{a+i}:=x_{b+i}$.
	Likewise, $[x_a,x_{a+i}]:=0$ denotes a sequence of adjacent
	edges labeled with $x_a:=0$, $x_{a+1}:=0$, $\dots$, $x_{a+i}:=0$.
	
	\begin{figure}[t]
		\centering
		\begin{tikzpicture}[shorten >=0pt,node distance=2cm,on grid,>=stealth',every state/.style={inner sep=0pt, minimum size=5mm, draw=blue!50,very thick,fill=blue!10}]
			\node[state]         (s)	{{$s$}}; 
			\node[state]         (sa) [below left=of s] {{$s_a$}};
			\node[state]         (sb) [below right=of s] {{$s_b$}};
			\node[state]         (ea) [below=of sa] {{$e_a$}};
			\node[state]         (eb) [below=of sb] {{$e_b$}};
			\node[state]         (e) [below right=of ea] {{$e$}};
			\node [draw, blue, fit=(sa)(ea), inner sep=7pt] {{}};
			\node [draw, red, fit=(sb)(eb), inner sep=7pt] {{}};
			\path[->]
			(s) edge [left] node  {{$x_1=c_1?$}} (sa)
			(s) edge [right] node  {{$\,x_1=c_2?$}} (sb)
			(sa) edge [left,dashed] node  {{$G_a\hspace*{-2.5pt}$}} (ea)
			(sb) edge [right,dashed] node  {{$\hspace*{-2.5pt}G_b$}} (eb)
			(ea) edge [right] node  {{}} (e)
			(eb) edge [right] node  {{}} (e);
		\end{tikzpicture}
		\caption{The BACFG $G_{r(a,b,c_1,c_2)}$,  
			output of the function $r$.}
		\label{fig:branching}
	\end{figure}
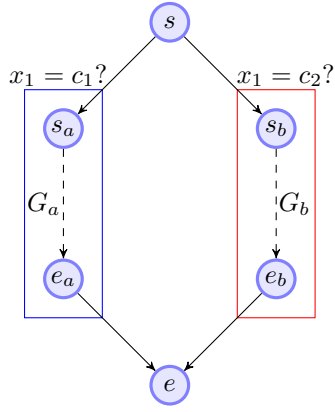

	\begin{proposition}\label{prop:branching}
		The state semantics of BACFGs in Definition \ref{def:cfg-relational-semantics} is branching.
	\end{proposition}
	\begin{proof}
		Let $\langle Q,=\rangle$ be the state semantics 
		of Definition \ref{def:cfg-relational-semantics}
		for a given predicate $Q\subseteq\wp(\mathbb{N}^t)$ on sets of states with $t\in\mathbb{N}$ variables.
		We define a total computable function $r:\mathbb{N}^4\to\mathbb{N}$ as follows: given two indices $a,b$ of BACFGs, say $G_a=( N_a,E_a,s_a,e_a)$ and $G_b=( N_b,E_b,s_b,e_b)$, and two values $c_1, c_2\in \mathbb{N}$, the function $r$ suitably renames the nodes of $G_a$ and $G_b$ to avoid clashes, and adds two fresh nodes
		$s$ (for start) and $e$ (for end) whose in/outgoing 
		edges are depicted by the BACFG in Fig.~\ref{fig:branching}, thus denoted by 
		$G_{r(a,b,c_1,c_2)}$.  
		%\todo{P: nella figura c'era $x_1=c_1?$, $x_1=c_2?$, sostituito con $y_1=c_1?$, $y_1=c_2?$, ok? L: no, perché la trasformazione non deve dipendere da $\vec{y}$; la variabile $x_1$ comunque conterrà esattamente il valore $y_1$}

		Observe that in the BACFG $G_{r(a,b,c_1,c_2)}$ with start and end nodes, resp., $s$ and $e$, with 
		inputs ranging in $\mathbb{N}^n$, for some $n\in \mathbb{N}$, the maximum variable index is $k=\max (k_a,k_b, n)$, where $k_a,k_b$ are, resp., the maximum variable indices in $G_a$ and $G_b$. Moreover, 
		for all inputs 
		$\Vec{y}=(y_1,y_2,\dots,y_n)\in\mathbb{N}^n$ and
		$c_1\neq c_2$, it turns out that:
		\begin{itemize}
			\item if $y_1=c_1$ then $\llbracket G_{r(a,b,c_1,c_2)} \rrbracket_{\Vec{y}}[e_a]=\llbracket G_a \rrbracket_{\Vec{y}} [e_a]\restriction_{k}$ and $\llbracket G_{r(a,b,c_1,c_2)} \rrbracket_{\Vec{y}}[e_b]=\varnothing$;
			\item if $y_1=c_2$ then $\llbracket G_{r(a,b,c_1,c_2)} \rrbracket_{\Vec{y}}[e_b]=\llbracket G_b\rrbracket_{\Vec{y}} [e_b]\restriction_{k}$ and $\llbracket G_{r(a,b,c_1,c_2)} \rrbracket_{\Vec{y}}[e_a]=\varnothing$;
			\item otherwise, i.e.\ when $y_1\notin\{c_1,c_2\}$, we have that $\llbracket G_{r(a,b,c_1,c_2)} \rrbracket_{\Vec{y}}[e_a]=\llbracket G_{r(a,b,c_1,c_2)} \rrbracket_{\Vec{y}}[e_b]= \varnothing$.
		\end{itemize}
		Consequently:
		{
			\begin{multline*}
			%\begin{split}
			\llbracket G_{r(a,b,c_1,c_2)} \rrbracket_{\Vec{y}}[e]
			%&
			=
			\llbracket G_{r(a,b,c_1,c_2)} \rrbracket_{\Vec{y}}[e_a]
			\cup
			\llbracket G_{r(a,b,c_1,c_2)} \rrbracket_{\Vec{y}}[e_b]
			%&
			=\\
			\begin{cases}
				\llbracket G_a \rrbracket_{\Vec{y}} [e_a]\restriction_{k}& \!\text{ if } y_1=c_1\\
				\llbracket G_b\rrbracket_{\Vec{y}} [e_b]\restriction_{k}& \!\text{ if } y_1=c_2\\
				\varnothing & \!\text{ otherwise.}
			\end{cases}
			%\end{split}
			\end{multline*}
		}
		Hence, $r$ is a total computable function such that for all $a,b,c_1,c_2,x\in\mathbb{N}$ with $c_1\neq c_2$:
		\begin{align*}
			&\lambda \Vec{y}. Q_{r(a,b,c_1,c_2)}^{(n)}(x,\Vec{y})\\
			&=
			\lambda \Vec{y}.
			\begin{cases}
				1& \text{ if } \llbracket G_{r(a,b,c_1,c_2)}\rrbracket_{x:\vec{y}}[e] \neq \varnothing \land \llbracket G_{r(a,b,c_1,c_2)}\rrbracket_{x:\vec{y}}[e]\restriction_{t}\in Q\\
				0 & \text{ if } \llbracket G_{r(a,b,c_1,c_2)}\rrbracket_{x:\vec{y}}[e] \neq \varnothing \land \llbracket G_{r(a,b,c_1,c_2)}\rrbracket_{x:\vec{y}}[e]\restriction_{t}\notin Q\\
				\uparrow & \text{ if } \llbracket G_{r(a,b,c_1,c_2)}\rrbracket_{x:\vec{y}}[e] = \varnothing
			\end{cases}\\
			&=
			\lambda \Vec{y}.
			\begin{cases}
				1& \text{ if } \llbracket G_a \rrbracket_{x:\vec{y}} [e_a]\neq \varnothing \land  \llbracket G_a \rrbracket_{x:\vec{y}} [e_a]\restriction_{k}\restriction_{t}\in Q \land x=c_1\\
				0& \text{ if } \llbracket G_a \rrbracket_{x:\vec{y}} [e_a]\neq \varnothing \land  \llbracket G_a\rrbracket_{x:\vec{y}} [e_a]\restriction_{k}\restriction_{t}\notin Q \land x=c_1\\
				\uparrow & \text{ if } \llbracket G_a \rrbracket_{x:\vec{y}} [e_a]= \varnothing \land x=c_1\\
				1& \text{ if } \llbracket G_b \rrbracket_{x:\vec{y}} [e_b]\neq \varnothing \land  \llbracket G_b\rrbracket_{x:\vec{y}} [e_b]\restriction_{k}\restriction_{t}\in Q \land x=c_2\\
				0& \text{ if } \llbracket G_b \rrbracket_{x:\vec{y}} [e_b]\neq \varnothing \land  \llbracket G_b\rrbracket_{x:\vec{y}} [e_b]\restriction_{k}\restriction_{t}\notin Q \land x=c_2\\
				\uparrow & \text{ if } \llbracket G_b \rrbracket_{x:\vec{y}} [e_b]= \varnothing\land x=c_2\\
				\uparrow & \text{ otherwise }
			\end{cases}\\
			&=
			\lambda \Vec{y}.
			\begin{cases}
				1& \text{ if } \llbracket G_a \rrbracket_{x:\vec{y}} [e_a]\neq \varnothing \land  \llbracket G_a \rrbracket_{x:\vec{y}} [e_a]\restriction_{t}\in Q \land x=c_1\\
				0& \text{ if } \llbracket G_a \rrbracket_{x:\vec{y}} [e_a]\neq \varnothing \land  \llbracket G_a\rrbracket_{x:\vec{y}} [e_a]\restriction_{t}\notin Q \land x=c_1\\
				\uparrow & \text{ if } \llbracket G_a \rrbracket_{x:\vec{y}} [e_a]= \varnothing\land x=c_1\\
				1& \text{ if } \llbracket G_b \rrbracket_{x:\vec{y}} [e_b]\neq \varnothing \land  \llbracket G_b\rrbracket_{x:\vec{y}} [e_b]\restriction_{t}\in Q \land x=c_2\\
				0& \text{ if } \llbracket G_b \rrbracket_{x:\vec{y}} [e_b]\neq \varnothing \land  \llbracket G_b\rrbracket_{x:\vec{y}} [e_b]\restriction_{t}\notin Q \land x=c_2\\
				\uparrow & \text{ if } \llbracket G_b \rrbracket_{x:\vec{y}} [e_b]= \varnothing\land x=c_2\\
				\uparrow & \text{ otherwise }
			\end{cases}\\
			&=
			\lambda \Vec{y}.
			\begin{cases}
				Q_a^{(n)}(x,\vec{y}) & \text{ if } x=c_1\\
				Q_b^{(n)}(x,\vec{y}) & \text{ if } x=c_2\\
				\uparrow & \text{ otherwise }
			\end{cases}\\
			&=
			\begin{cases}
				\lambda \Vec{y}.Q_a^{(n)}(x,\vec{y}) & \text{ if } x=c_1\\
				\lambda \Vec{y}.Q_b^{(n)}(x,\vec{y}) & \text{ if } x=c_2\\
				\lambda \Vec{y}.\uparrow & \text{ otherwise }
			\end{cases}
		\end{align*}
		Therefore, $r$ satisfies the branching property of
		Definition~\ref{de:branching}.
	\end{proof}

	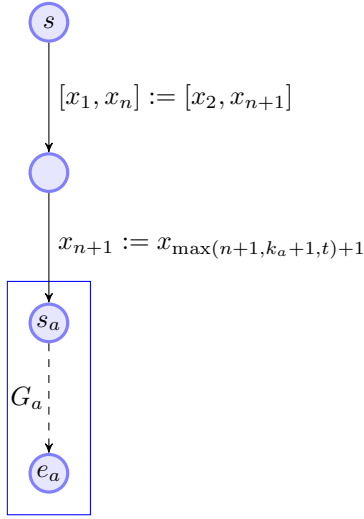
\begin{figure}
		\centering
		\begin{tikzpicture}[shorten >=0pt,node distance=2cm,on grid,>=stealth',every state/.style={inner sep=0pt, minimum size=5mm, draw=blue!50,very thick,fill=blue!10}]
			\node[state]         (s)	{{$s$}}; 
			\node[state]         (q1)[below=of s]	{{}}; 
			\node[state]         (sa)[below=of q1]	{{$s_a$}}; 
			\node[state]         (ea) [below=of sa] {{$e_a$}};
			\node [draw, blue, fit=(sa)(ea), inner sep=8pt] {{}};
			\path[->]
			(s) edge [right] node  {{$[x_{1},x_{n}]:=[x_2,x_{n+1}]$}} (q1)
			(q1) edge [right] node  {{$x_{n+1}:=x_{\max(n+1,k_a+1,t)+1}$}} (sa)
			(sa) edge [left,dashed] node  {{$G_{a}\!$}} (ea);
		\end{tikzpicture}
		\caption{The BACFG $G_{r(a)}$,  
			output of the function $r$, where irrelevant node names are omitted.}
		\label{fig:discharging}
	\end{figure}
	
	\begin{proposition}
		The state semantics of  BACFGs in Definition \ref{def:cfg-relational-semantics} is discharging.
	\end{proposition}
	\begin{proof}
		Let $\langle Q,=\rangle$ be a state semantics for a predicate $Q\subseteq\wp(\mathbb{N}^t)$ on sets of states with $t\in\mathbb{N}$ variables.
		Similarly to the proof of Proposition~\ref{prop:branching},
		let us define a total computable function $r:\mathbb{N}\to\mathbb{N}$ as follows: given an index $a$ of a BACFG $G_a=( N_a,E_a,s_a,e_a)$, where $k_a$ is the maximum variable index occuring in $G_a$, the function $r$ suitably renames the nodes of $G_a$ to avoid clashes, and adds two fresh nodes
		$s$ and $e$ whose in/outgoing edges are depicted by the BACFG in Fig.~\ref{fig:discharging}.
		
		Notice that in the BACFG $G_{r(a)}$ with start and end nodes, resp., $s$ and $e_a$, given $n\geq 1$, for all input $\Vec{y}=(y_1,y_2,\dots,y_n)\in\mathbb{N}^n$ and $x\in\mathbb{N}$ we have that $\llbracket G_{r(a)}\rrbracket_{x:\vec{y}} [e_a]\!\restriction_t=\llbracket G_a \rrbracket_{\Vec{y}} [e_a]\!\restriction_{t}$: this happens because the command $[x_{1},x_{n}]:=[x_2,x_{n+1}]$ left shifts the variables and the assignment $x_{n+1}:=x_{\max(n+1,k_a+1,t)+1}$ guarantees that $x_{n+1}$ is undefined.
		Hence, $r$ is a total computable function such that for all $a,x\in\mathbb{N}$:
		\begin{align*}
			&\lambda \Vec{y}. Q_{r(a)}^{(n+1)}(x,\Vec{y})\\
			&=\lambda \Vec{y}.
			\begin{cases}
				1& \text{ if } \llbracket G_{r(a)}\rrbracket_{x:\vec{y}}[e_a] \neq \varnothing \land \llbracket G_{r(a)}\rrbracket_{x:\vec{y}}[e_a]\restriction_{t}\in Q\\
				0 & \text{ if } \llbracket G_{r(a)}\rrbracket_{x:\vec{y}}[e_a] \neq \varnothing \land \llbracket G_{r(a)}\rrbracket_{x:\vec{y}}[e_a]\restriction_{t}\notin Q\\
				\uparrow & \text{ if } \llbracket G_{r(a)}\rrbracket_{x:\vec{y}}[e_a] = \varnothing
			\end{cases}\\
			&=\lambda \Vec{y}.
			\begin{cases}
				1& \text{ if } \llbracket G_{a}\rrbracket_{\vec{y}}[e_a] \neq \varnothing \land \llbracket G_{a}\rrbracket_{\vec{y}}[e_a]\restriction_{t}\in Q\\
				0 & \text{ if } \llbracket G_{a}\rrbracket_{\vec{y}}[e_a] \neq \varnothing \land \llbracket G_{a}\rrbracket_{\vec{y}}[e_a]\restriction_{t}\notin Q\\
				\uparrow & \text{ if } \llbracket G_{a}\rrbracket_{\vec{y}}[e_a] = \varnothing
			\end{cases}\\
			&= \lambda \Vec{y}.Q_{a}^{(n)}(\vec{y}).
		\end{align*}
		Thus, $r$ is a function satisfying the discharging property of Definition~\ref{de:branching}.
	\end{proof}

	\begin{proposition}
		The state semantics of BACFGs in Definition \ref{def:cfg-relational-semantics} is ssmn.
	\end{proposition}
	\begin{proof}
		Let $m,n\geq 1$ and $\langle Q,=\rangle$ be a state semantics for a given predicate $Q\subseteq\wp(\mathbb{N}^p)$ on sets of states with $p\in\mathbb{N}$ variables. We define a total computable function $s:\mathbb{N}^{m+2}\to\mathbb{N}$ which takes as input two indices $a,b$ and a $m$-dimensional vector $\Vec{z}\in\mathbb{N}^m$. Intuitively, to satisfy the ssmn property of Definition~\ref{def:smnsemantics}, the output of $s(a,b,\Vec{z})$ should be a BACFG that simulates the computation of the concrete semantics $\phi_b^{(m)}$ as defined in 
		\eqref{eq:urm-concrete-semantics}. Since this latter concrete semantics is defined on URMs, it is enough to simulate the program $RM_b=(I_1,\dots,I_t)$. To this aim, recall that the total computable function $\tau$ of Definition~\ref{def:tau} transforms URMs into BACFGs having equivalent semantics. Roughly, the BACFG $G_{s(a,b,\Vec{z})}$ on input $\vec{y}\in\mathbb{N}^n$ first simulates $G_{b'}=(N_{b'},E_{b'},q_1,q_{t+1})\in \tau (RM_b)$ on input $\Vec{z}$, and, then, simulates $G_a=(N_a,E_a,s_a,e_a)$ on input $\phi_b^{(m)}(\vec{z}):\vec{y}$.
		Before going into the details, recall that, in general, URMs set unused registers to $0$, so that, by a slight abuse of notation, we define the vector projection $\vec{z}\!\restriction_{k}\in \mathbb{N}^{k}$, for all $\vec{z}=(z_1,\dots, z_{k'})\in \mathbb{N}^{k'}$, as follows:
		\[
		\vec{z}\!\restriction_{k}\:\eqd
		\begin{cases}
			(z_1,\dots,z_{k'},0)\!\restriction_{k} & \text{ if } 0 \leq k' < k\\
			(z_1,\dots,z_k)& \text{ if } k\leq k'\\
		\end{cases}
		\]
		Let $k_{a}$ and $k_b$ be the maximum variable (or register) index occuring, resp., in $G_a$ and $RM_{b}$. Recall the operational semantics $\Rightarrow$ for URMs of Definition~\ref{de:operational-semantics-URM} and notice that:
		\[\phi_b^{(m)}(\vec{z})=
		\begin{cases}
			z'_1 & \text{ if } \exists \vec{z}'\in\mathbb{N}^{k_b}.\langle \vec{z}\restriction_{k_b}, 1 \rangle \Rightarrow^* \langle \vec{z}', t+1 \rangle\\
			\uparrow  & \text{ otherwise.}
		\end{cases}
		\]
		Therefore, by Propositions~\ref{prop:urm->cfg-halt} and~\ref{prop:urm->cfg-diverge}, in order to simulate $\phi_b^{(m)}(\vec{z})$ it is enough to execute $G_{b'}$ on input $\vec{z}\restriction_{k_b}:0$. More in detail, the transform $s(a,b,\Vec{z})$ will add the following commands:
		
		\begin{enumerate}
			\item $[x_{k_b+2},x_{k_b+n+1}]:=[x_1,x_n]$, to safely store the original input $\vec{y}\in\mathbb{N}^n$; in fact, the execution of $\llbracket G_{b'}\rrbracket_{\vec{z}\restriction_{k_b}:0}$ will use the first $k_b+1$ variables only;
			\item $[x_1,x_{\min (m, k_b)}]:=[z_1,z_{\min (m, k_b)}]$, so that the first $\min(m, k_b)$ variables contain $\vec{z}\restriction_{k_b}$ except for the $0$-padding;
			\item $[x_{\min (m, k_b)+1}, x_{k_b+1}]:=0$, to (possibly) add the missing $0$-padding;
		\end{enumerate}

		This allows us to execute $G_{b'}$ on input $(\vec{z}\restriction_{k_b}:0):\vec{y}$. The next step is to execute $G_a$ on input $\phi_b^{(m)}(\vec{z}):\vec{y}$. Therefore, we add the following commands:
		
		\begin{enumerate}\setcounter{enumi}{3}
			\item $[x_2,x_{n+1}]:=[x_{k_b+2},x_{k_b+n+1}]$, to restore the original input ($\vec{y}$) on the variables starting from $x_2$;
			\item $[x_{n+2},x_{\max(k_a,p)}]:=[x_{k_b+n+2},x_{k_b+\max(k_a,p)}]$, to ensure that all the remaining variables up to $x_{\max(k_a,p)}$ are left undefined.
		\end{enumerate}

		Finally, the BACFG $G_a$ is executed. The resulting BACFG $G_{s(a,b,\Vec{z})}$, with start and end nodes  $s$ and $e_a$, resp., is described by the graph in Fig.~\ref{fig:smn}.
		Observe that, by definition:
		
		\begin{itemize}
			\item if $\phi_b^{(m)}(\Vec{z})\uparrow$ then, by Proposition~\ref{prop:urm->cfg-diverge}, $\llbracket G_{s(a,b,\Vec{z})} \rrbracket_{\Vec{y}}[e_a]=\llbracket G_{b'} \rrbracket_{\Vec{z}}[q_{t+1}]=\varnothing$;
			\item otherwise, by Proposition~\ref{prop:urm->cfg-halt}, $\llbracket G_{s(a,b,\Vec{z})} \rrbracket_{\Vec{y}}[e_a]\!\restriction_p \, =\llbracket G_a \rrbracket_{\phi_b^{(m)}(\Vec{z}):\Vec{y}} [e_a]\!\restriction_{p}$.
		\end{itemize}

		Hence, we defined a total computable function $s$ such that for all $a,b\in\mathbb{N}$ and $\Vec{z}\in\mathbb{N}^m$:
		\begin{align*}
			&\lambda \Vec{y}. Q_{s(a,b,\Vec{z})}^{(n)}(\Vec{y})\\
			&=
			\lambda \Vec{y}.
			\begin{cases}
				1& \text{ if } \llbracket G_{s(a,b,\Vec{z})}\rrbracket_{\vec{y}}[e_a] \neq \varnothing \land \llbracket G_{s(a,b,\Vec{z})}\rrbracket_{\vec{y}}[e_a]\restriction_{p}\in Q\\
				0& \text{ if } \llbracket G_{s(a,b,\Vec{z})}\rrbracket_{\vec{y}}[e_a] \neq \varnothing \land \llbracket G_{s(a,b,\Vec{z})}\rrbracket_{\vec{y}}[e_a]\restriction_{p}\notin Q\\
				\uparrow & \text{ if } \llbracket G_{s(a,b,\vec{z})}\rrbracket_{\vec{y}}[e_a] = \varnothing
			\end{cases}\\
			&=
			\lambda \Vec{y}.
			\begin{cases}
				1& \text{ if } \phi_b^{(m)}(\Vec{z})\downarrow \land  \llbracket G_a \rrbracket_{\phi_b^{(m)}(\Vec{z}):\Vec{y}} [e_a]\neq \varnothing \land \llbracket G_a \rrbracket_{\phi_b^{(m)}(\Vec{z}):\Vec{y}} [e_a]\restriction_{p}\in Q\\
				0& \text{ if }  \phi_b^{(m)}(\Vec{z})\downarrow \land \llbracket G_a \rrbracket_{\phi_b^{(m)}(\Vec{z}):\Vec{y}} [e_a] \neq \varnothing \land \llbracket G_a \rrbracket_{\phi_b^{(m)}(\Vec{z}):\Vec{y}} [e_a]\restriction_{p}\notin Q\\
				\uparrow & \text{ otherwise }
			\end{cases}\\
			&=
			\lambda \Vec{y}.Q_a^{(n+1)}(\phi_b^{(m)}(\Vec{z}), \Vec{y})
		\end{align*}
		Hence, $s$ is a function satisfying the ssmn property of Definition~\ref{def:smnsemantics}.
	\end{proof}
	
	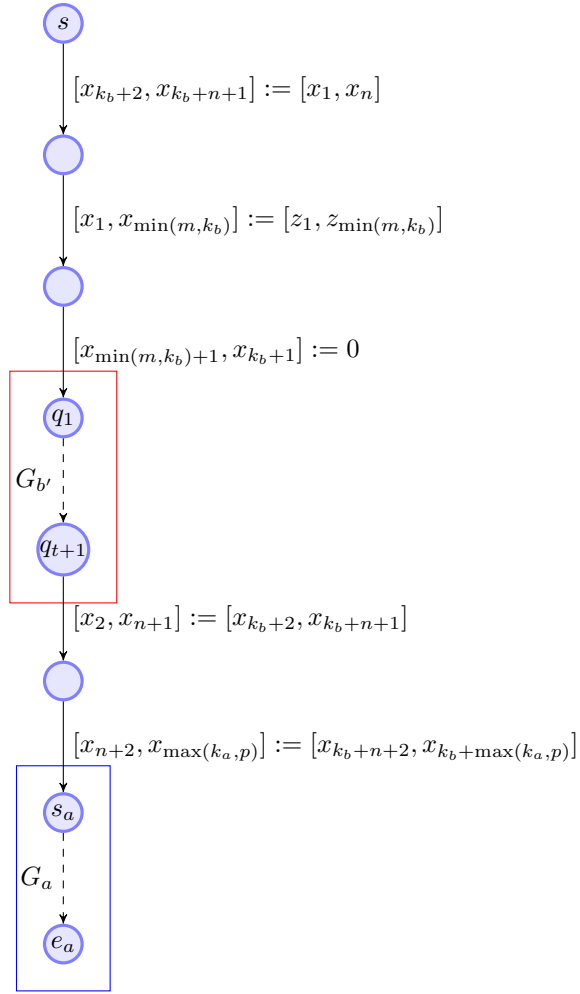
\begin{figure}
		\centering
		\begin{tikzpicture}[shorten >=0pt,node distance=1.75cm,on grid,>=stealth',every state/.style={inner sep=0pt, minimum size=5mm, draw=blue!50,very thick,fill=blue!10}]
			\node[state]         (s)	{{$s$}}; 
			\node[state]         (q1) [below=of s] {{}};
			\node[state]         (q2) [below=of q1] {{}};
			\node[state]         (sb) [below=of q2] {{$q_1$}};
			\node[state]         (eb) [below=of sb] {{$q_{t+1}$}};
			\node[state]         (q3) [below=of eb] {{}};
			\node[state]         (sa) [below=of q3] {{$s_a$}};
			\node[state]         (ea) [below=of sa] {{$e_a$}};
			\node [draw, blue, fit=(sa)(ea), inner sep=10pt] {{}};
			\node [draw, red, fit=(sb)(eb), inner sep=10pt] {{}};
			\path[->]
			(s) edge [right] node  {{$[x_{k_b+2},x_{k_b+n+1}]:=[x_1,x_n]$}} (q1)
			(q1) edge [right] node  {{$[x_1,x_{\min (m, k_b)}]:=[z_1,z_{\min (m, k_b)}]$}} (q2)
			(q2) edge [right] node  {{$[x_{\min (m, k_b)+1}, x_{k_b+1}]:=0$}} (sb)
			(sb) edge [left, dashed] node  {{$G_{b'}$}} (eb)
			(eb) edge [right] node  {{$[x_2,x_{n+1}]:=[x_{k_b+2},x_{k_b+n+1}]$}} (q3)
			(q3) edge [right] node  {{$[x_{n+2},x_{\max(k_a,p)}]:=[x_{k_b+n+2},x_{k_b+\max(k_a,p)}]$}} (sa)
			(sa) edge [left, dashed] node  {{$G_a$}} (ea);
		\end{tikzpicture}
		\caption{The BACFG $G_{s(a,b,\vec{z})}$,  
			output of the function $s$, where irrelevant node names are omitted.}
		\label{fig:smn}
	\end{figure}
	
	\subsection{An Application to Affine Program Invariants}
	
	Consider a state semantics 
	$\langle {Q}, =\rangle$
	for some predicate $Q \subseteq \wp(\mathbb{N}^t)$.  For all $n\geq 1$, let us define
	two sets  $A^{\forall{Q}}$ and $A^{\exists{Q}}$, 
	by distinguishing two cases depending on whether $Q$ 
	includes the empty set, that models nontermination, or not: 
	
	\begin{enumerate}[(1)]
		\item If $\varnothing \notin Q$ then:
		\begin{itemize}
			\item[]
			$A^{\forall{Q}} \eqd \{a \in \mathbb{N} \mid  \forall \Vec{y}.\, {Q}_a^{(n)}(\Vec{y})=1\}$;
			\item[]
			$A^{\exists{Q}}\eqd\{ a \in\mathbb{N} \mid \exists \Vec{y}.\, {Q}_a^{(n)}(\Vec{y})=1\}$.
		\end{itemize} 
		\item
		If $\varnothing \in Q$ then: 
		\begin{itemize}
			\item[] $A^{\forall{Q}} \eqd \{a \in \mathbb{N} \mid  \forall \Vec{y}.\, {Q}_a^{(n)}(\Vec{y})\in\{1,\uparrow\}\}$;
			\item[] 
			$A^{\exists{Q}}\eqd\{ a \in\mathbb{N} \mid \exists \Vec{y}.\, {Q}_a^{(n)}(\Vec{y})\in\{1,\uparrow\}\}$.
		\end{itemize}
	\end{enumerate}
	
	Hence, $A^{\forall{Q}}$ ($A^{\exists{Q}}$) is 
	the set of BACFGs such that $Q$ holds at $e_a$ for any (some) input state.
	It turns out that if the property $Q$  is nontrivial then neither $A^{\forall{Q}}$ nor $A^{\exists{Q}}$ can
	be recursive.
	% Indeed, observe that $A^{\forall{Q}}$ is
	% $\sim_{{Q}}$-extensional,
	% so that Theorem~\ref{th:state-semantics-properties} enables applying 
	% Theorem~\ref{th:rice-branching}
	% to $\langle {Q}, =\rangle$. The same argument applies to the existential version $A^{\exists{Q}}$. We have therefore the following consequence. 
	
	\begin{corollary}\label{coro:affine} 
		If $Q$ is not trivial then 
		$A^{\forall{Q}}$ and $A^{\exists{Q}}$ are not recursive. 
	\end{corollary}
	\begin{proof}
		Observe that $A^{\forall{Q}}$ is $\sim_{{Q}}$-extensional. Thus,
		Theorem~\ref{th:state-semantics-properties} enables applying our intensional
		Theorem~\ref{th:rice-branching} to the state semantics $\langle {Q}, =\rangle$ 
		to derive that $A^{\forall{Q}}$ is not recursive. 
		The same argument applies to the existential version
		$A^{\exists{Q}}$.
		%\todo{P: era un commento externo, trasformato in prova}
	\end{proof}
	
	Thus, Corollary~\ref{coro:affine} means that we cannot decide if a
	nontrivial predicate $Q$ holds at a given program point of a
	BACFG for all input states, neither whether there exists an input state that will
	make $Q$ true.
	Let us remark that the predicates $Q$ are arbitrary and
	include, but are not limited to, relational predicates between
	program variables such as affine equalities of Karr's abstract domain.
	Let us define some noteworthy examples of predicates:
	
	\begin{enumerate}[(i)]
		\item Given a set of affine equalities $\mathit{aff} =\{\vec{a_j}\cdot \vec{x} = b_j \}_{j=1}^m$, with    
		$\vec{a_j}\in \mathbb{Z}^t$ and $b_j\in \mathbb{Z}$,  
		$Q_{\mathit{aff}}\eqd \{S\in\wp(\mathbb{N}^t) \mid \forall \vec{v}\in S.
		\forall j\in [1,m].\:  \vec{a_j}\cdot \vec{v}=b_j\}$;
		\item Given $i\in [1,t]$ and $c\in \mathbb{N}$, $Q_{=c}\eqd \{S\in\wp(\mathbb{N}^t) \mid \exists \vec{v}\in S.\ v_i= c\}$;
		\item Given a size $k\in \mathbb{N}$, $Q_{\text{fin}_k}\eqd 
		\{S\in\wp(\mathbb{N}^t) \mid |S|=k\}$ and
		$Q_{\text{fin}}\eqd \cup_{k\in \mathbb{N}} Q_{\text{fin}_k}$. 
	\end{enumerate}

	Therefore, it turns out that Corollary~\ref{coro:affine} for
	$A^{\forall Q_{\mathit{aff}}}$ entails the undecidability
	result of M\"uller-Olm and Seidl~\cite[Section~7]{MOS04Affine} discussed
	at the beginning of Section~\ref{se:cfg}.  
	The predicate $Q_{=c}$ can be used to derive 
	the undecidability  
	of checking if some variable $x_i$ may store a given constant $c$
	for affine programs with positive affine guards, e.g., for $c=0$ this amounts
	to the undecidability of detecting division-by-zero bugs.  
	Finally, with $Q_{\text{fin}_0}$ we obtain the undecidability of dead code elimination,
	$Q_{\text{fin}_1}$ entails the well-known undecidability of constant detection \cite{hecht,reif77},
	while the existential predicate $Q_{\text{fin}}$ encodes whether 
	some program point may only have finitely many different states. 
	
	\section{Discussion of Related Work}
	\label{se:related}
	In this section we discuss in detail the relation with some of Asperti's results \cite{Asp08} and with Rogers' systems of indices \cite{Rog58,Rog67}. 
	
	\subsection{Relation with Asperti's Approach}
	\label{subse:asperti}
	
	We show that our ssmn property in Definition~\ref{def:smnsemantics} is a generalisation of the smn property in Asperti's approach~\cite{Asp08}, in a way that  Kleene's second recursion theorem and Rice's theorem for complexity cliques in~\cite{Asp08} arise as instances of the corresponding results in our approach.
	Let us first recall and elaborate on the axioms for the complexity of function composition studied by Lischke~\cite{Lischke75,Lischke76,Lischke77} and assumed in~\cite[Section 4]{Asp08}. 
	
	\begin{definition}[{\bf Linear time and space complexity composition}]\label{def-ltsc}
		Consider a given concrete semantics $\phi$ and a Blum complexity $\Phi$. 
		The pair $\langle \phi, \Phi \rangle$ has the \emph{linear time
			composition} property if there exists a total computable function
		$h:\mathbb{N}^2\rightarrow\mathbb{N}$ such that for all $i,j\in\mathbb{N}$:
		
		\begin{enumerate}[{\rm (1)}]
			\item $\phi_{h(i,j)}=\phi_i \circ \phi_j$,
			\item $\Phi_{h(i,j)}\in\Theta(\Phi_i \circ \phi_j + \Phi_j)$. 
		\end{enumerate}

		If (2) is replaced by 
		
		\begin{enumerate}[{\rm (2$'$)}]
			\item $\Phi_{h(i,j)}\in\Theta(\max\{\Phi_i \circ \phi_j, \Phi_j\})$ 
		\end{enumerate}
		then $\langle \phi, \Phi \rangle$ is said to have the \emph{linear
			space composition} property.
		\qed
	\end{definition}
	
	Roughly speaking, the linear time composition property states that
	there exists a program $h(i,j)$ which computes the composition
	$\phi_i(\phi_j(x))$ in an amount of time which is asymptotically
	equivalent to the sum of the time needed for computing $P_j$ on input $x$, eventually producing some output $\phi_j(x)$, and the time form computing $P_i$ on such value.  On the other
	hand, the linear space composition property aims at modeling the
	needed space, so that rather than adding the complexities of $P_i$ and
	$P_j$, their maximum is considered, since this intuitively is the
	maximum amount of space needed for computing a composition of
	programs.
	
	By observing that
	$\Theta(\max\{\Phi_i \circ \phi_j, \Phi_j\})=\Theta(\Phi_i \circ
	\phi_j+\Phi_j)$ we can merge the linear time and space properties of 
	Definition~\ref{def-ltsc}  and extend them
	for $n$-ary compositions as follows. 
	
	\begin{definition}[{\bf Linear complexity composition}]
		\label{def:linearcomp}
		Given a concrete semantics $\phi$ and a Blum complexity $\Phi$, 
		the pair $\langle \phi, \Phi \rangle$ has the \emph{linear complexity
			composition} property if, given $n,m\geq 1$, there exists a
		total computable function $h:\mathbb{N}^2\rightarrow\mathbb{N}$ such
		that for all $i,j\in\mathbb{N}$:
		
		\begin{itemize}
			\item[] \(\phi_{h(i,j)}^{(m+n)}=\lambda \vec{x}\lambda \vec{y}.\:\phi_i^{(n+1)}(\phi_j^{(m)}(\vec{x}), \Vec{y})\),
			\item[] \(\Phi_{h(i,j)}^{(m+n)}\in\Theta(\lambda \vec{x}\lambda \vec{y}.\:(
			\Phi_i^{(n+1)}(\phi_j^{(m)}(\vec{x}),\vec{y}))
			+\Phi_j^{(m)}(\vec{x})
			))   		
			\). \qed
		\end{itemize}
	\end{definition}

	We can now recall the smn property as defined in~\cite[Definition~11]{Asp08}.
	
	\begin{definition}[{\bf Asperti's smn property}]
		\label{def:aspertismn}
		Given a concrete semantics $\phi$, a Blum
		complexity $\Phi$ and  $m,n\geq 1$, the pair $\langle \phi, \Phi \rangle$ has the
		\emph{Asperti's smn} property if there exists a total computable
		function $s:\mathbb{N}^{m+1}\rightarrow\mathbb{N}$ such that
		$\forall e\in\mathbb{N},\Vec{x}\in\mathbb{N}^m$:
		
		\begin{itemize}
			\item[] $\lambda \Vec{y}.\phi_e^{(m+n)}(\Vec{x}, \Vec{y}) = \phi_{s(e, \Vec{x})}^{(n)}$,
			\item[] $\lambda \Vec{y}.\Phi_e^{(m+n)}(\Vec{x}, \Vec{y}) \in \Theta(\lambda \Vec{y}.\Phi_{s(e, \Vec{x})}^{(n)}(\Vec{y}))$. \qed
		\end{itemize}
	\end{definition}
	
	Informally, the smn property of Definition~\ref{def:aspertismn} states
	that the operation of fixing parameters preserves both the concrete
	semantics and the asymptotic complexity. Under these assumptions, we
	can show that Asperti's complexity clique semantics satisfies our ssmn
	property. The proof is a simple adaptation of the one used
	in Section~\ref{se:smn-fair} to argue that the concrete semantics of
	Example~\ref{ex:concrete} is ssmn.
	
	\begin{lemma}\label{prop:smn-compl-clique}
		Let $\langle \pi, \equiv_\pi\rangle$ be the complexity clique
		semantics of Example \ref{ex:asperti}. If $\langle \pi, \equiv_\pi \rangle$
		satisfies Asperti's smn and linear complexity composition properties
		then $\langle \pi, \equiv_\pi\rangle$ is ssmn.
	\end{lemma}
	
	\begin{proof}
		We have to show that given $m,n \geq 1$, there exists a total
		computable function $s : \mathbb{N}^{m+2}\to \mathbb{N}$ such that
		for all $a,b \in \mathbb{N}$, $\vec{x} \in \mathbb{N}^m$:
		\begin{equation*}
			\lambda \vec{y}. \pi_a^{(n+1)}(\phi_b^{(m)}(\vec{x}), \Vec{y})
			\equiv_\pi
			\pi_{s(a,b,\vec{x})}^{(n)}.
		\end{equation*}
		
		We have that
		\begin{align*}
			& \lambda \vec{y}. \pi_a^{(n+1)}(\phi_b^{(m)}(\vec{x}), \Vec{y}) = &\\
			& \quad = \lambda \vec{y}. \pair{\phi_a^{(n+1)}(\phi_b^{(m)}(\vec{x}), \Vec{y})}{\Phi_a^{(n+1)}(\phi_b^{(m)}(\vec{x}), \Vec{y})}\\
			& \quad [\mbox{by definition of $\pi_a$}]\\
			& \quad \equiv_{\pi} \lambda \vec{y}. \pair{\phi^{(m+n)}_{h(a,b)}(\vec{x}, \Vec{y})}{\Phi_{h(a,b)}^{(m+n)}(\vec{x}, \Vec{y})}\\
			& \quad [\mbox{with $h:\mathbb{N}^2\rightarrow\mathbb{N}$ total computable, by linear complexity composition}]\\
			& \quad \equiv_{\pi} \lambda \vec{y}. \pair{\phi^{(n)}_{s'(h(a,b),\vec{x})}(\Vec{y})}{\Phi_{s'(h(a,b),\vec{x})}^{(n)}(\Vec{y})}\\
			& \quad [\mbox{with $s':\mathbb{N}^{m+1}\rightarrow\mathbb{N}$ total computable, by Asperti's smn property}]\\
			& \quad = \pair{\phi^{(n)}_{s'(h(a,b),\vec{x})}}{\Phi_{s'(h(a,b),\vec{x})}^{(n)}} = \pi_{s'(h(a,b),\vec{x})}^{(n)} 
		\end{align*}
		The desired function $s : \mathbb{N}^{m+2}\to \mathbb{N}$ can therefore
		be defined as
		$s(a,b,\vec{x}) \eqd s'(h(a,b),\vec{x})$. Note that
		$s$ is total computable since $h$ and $s'$ are so.
	\end{proof}
	
	This result, together with the observation that the notion of fairness (Definition~\ref{de:univ-prg}) instantiated to
	the complexity clique semantics is exactly that of~\cite[Definition 26]{Asp08}, allows us to retrieve Kleene's second recursion theorem and Rice's theorem for complexity cliques in~\cite{Asp08} 
	as instances of our corresponding results  in Section~\ref{se:kleene-rice}.

	\subsection{Relation with Systems of Indices}
	\label{subse:indices}
	
	As mentioned in Section~\ref{se:basics}, our definition of abstract semantics resembles the acceptable 
	systems of indices~\cite[Definition II.5.1]{Odi89} 
	or numberings~\cite[Exercise 2-10]{Rog67}, firstly studied by Rogers~\cite{Rog58}. In this section we discuss how such notions compare.
	
	\begin{definition}[{\bf System of indices~{\cite[Definition
			II.5.1]{Odi89}}}]\label{def-soi}
		A \emph{system of indices} is a family of functions
		$\{\psi^n\}_{n\in\mathbb{N}}$ such that each
		$\psi^n:\mathbb{N}\to\mathcal{C}_n$ is a surjective map that
		associates program indices to $n$-ary partial recursive functions.
		
		\begin{itemize}
			\item $\{\psi^n\}_{n\in\mathbb{N}}$ has the \emph{parametrization}
			(or smn) property if for every $m, n\in\mathbb{N}$ there is a
			total computable function $s:\mathbb{N}^{m+1}\to\mathbb{N}$ such
			that $\forall e\in\mathbb{N},\Vec{x}\in\mathbb{N}^m$:
			\[
			\lambda \vec{y}. \psi_e^{m+n}(\vec{x}, \Vec{y}) =
			\psi^{n}_{s(e,\vec{x})}.
			\]
			\item $\{\psi^n\}_{n\in\mathbb{N}}$ has the \emph{enumeration}
			property if for every $n\in\mathbb{N}$ there exists
			$u\in\mathbb{N}$ such that for all and $e\in\mathbb{N}$ and
			$\Vec{y}\in\mathbb{N}^n$:
			\[
			\psi_e^{n}=\lambda \vec{y}.\psi_u^{n+1}(e, \vec{y}).\tag*{\qed}
			\] 
		\end{itemize}
	\end{definition}

	Any standard Gödel numbering associating a program with the function
	it computes is a system of indices with the parametrization and
	enumeration properties. Moreover, exactly as we did in
	Example~\ref{ex:concrete}, any system of indices
	$\{\psi^n\}_{n\in\mathbb{N}}$ can be viewed as an abstract semantics
	$\langle \pi , =\rangle$ with $\pi^a_n\triangleq \psi_a^n$. In this context, the
	enumeration and parametrization properties correspond  to our fairness
	and ssmn conditions: fairness is exactly enumeration while ssmn
	follows from parametrization and enumeration, as discussed in Section~\ref{se:smn-fair} for
	the concrete semantics (cf.~Example~\ref{ex:concrete}).
	
	A system of indices is defined to be \emph{acceptable} if it allows to
	get back and forth with a given system of indices satisfying the
	parametrization and enumeration properties through a pair of total
	computable functions.
	
	\begin{definition}[{\bf Acceptable system of indices~{\cite[Definition~4]{Rog58}}}]\label{def-asoi}\rm
		Let $\{\varphi^n\}_{n\in\mathbb{N}}$ be a given system of indices
		with the parametrization and enumeration properties.  A system of
		indices $\{\psi^n\}_{n\in\mathbb{N}}$ is \emph{acceptable} if there
		exist two total computable functions $f,g:\mathbb{N}\to\mathbb{N}$
		such that for all $a,n\in\mathbb{N}$:
		\begin{equation*}
			\psi_a^n = \varphi_{f(a)}^n \quad \text{ and } \quad \varphi_a^n =\psi_{g(a)}^n. \tag*{\qed}
		\end{equation*}
	\end{definition}
	
	As shown in \cite[Proposition~II.5.3]{Odi89}, 
	it turns out that a system of indices is acceptable if and only if it satisfies both
	enumeration and parametrization (a proof of this characterization was first given by 
	Rogers~\cite[Section~2]{Rog58}). 
	Consequently, an acceptable system of indices $\{\psi^n\}_{n\in\mathbb{N}}$ can be viewed as an abstract
	semantic $\langle \pi, = \rangle$, where $\pi_a^n = \psi_a^n$, which, by this characterization of 
	acceptability, is 
	ssmn and fair, and therefore, by Theorem~\ref{th:second-recursion} it enjoys Kleene's second recursion theorem, as already known from~\cite[Corollary~II.5.4]{Odi89}.
	
	Under this perspective, a generic abstract semantics according to Definition~\ref{de:sem-fw} can be viewed as a proper generalisation
	of the notion of acceptable system of indices, in the sense that the latter merely encodes a change of program numbering and does not allow 
	to take into account an actual abstraction of the concrete input/output behaviour of programs.

	\section{Conclusion and Future Work}
	\label{se:conclusions}
	
	This work generalises some traditional extensional results of computability theory, notably 
	Kleene's second recursion theorem and Rice's theorem, to intensional abstract program semantics that include the complexity cliques investigated by Asperti~\cite{Asp08}. 
	Our approach was also inspired by Moyen and Simonsen~\cite{MS19}\ and relies on weakening the classical definition of extensional program property to a notion of partial extensionality w.r.t.\ abstract program semantics 
	that satisfy some structural conditions.
	As an application, we strengthened and generalised 
	a result by M\"{u}ller-Olm and Seidl~\cite{MOS04Affine} proving that
	for affine programs with positive affine
	guards it is undecidable 
	whether an affine relation holds at a given program point.
	Our results also shed further light on the claim that these undecidability results 
	hinge on the Turing completeness of the underlying computational model, as argued in~\cite{MS19}.
	
	As future work, a natural question would be to investigate intensional extensions of
	Rice-Shapiro's theorem that fit our framework based on abstract semantics. 
	This appears to be a nontrivial challenge. Generalisations of Rice-Shapiro's theorem have been given in \cite[Section~5]{Asp08} and \cite[Section~5.1]{MS19}. 
	A generalisation in the vein of the approach in~\cite{Asp08} seems
	to be viable, but would require structural assumptions on abstract program semantics
	that, while natural in~\cite{Asp08} whose focus is on complexity properties, would be artificial for
	abstract program semantics and would limit a general applicability.
	A further stimulating research topic is to apply our approach to abstract semantics as
	defined by abstract interpretation of programs~\cite{cousot21}, in particular for
	investigating the relationship with the notion of abstract
	extensionality studied by Bruni et al.~\cite{BGG+20}. Finally, while our framework 
	relies on the assumption of an underlying
	Turing complete computational model,  in a different direction, one
	could try to consider intensional properties for classes of programs
	indexing subrecursive functions (e.g., primitive recursive functions), 
	whose extensional properties have
	been already studied (see, e.g.,~\cite{Hoy:DPSF,Koz:ISC}). Despite the fact that we
	suppose that our approach will fall short on these program classes, as one
	cannot expect to have a universal program inside the class itself or the
	validity of Kleene's second recursion theorem, we think that 
	this represents an intriguing research challenge.
	
	\section*{Acknowledgements}
	We are grateful to Roberto Giacobazzi for thorough discussions and comments. 
	
	Paolo Baldan and Francesco Ranzato have been partially funded by \emph{University of Padova}, under the SID2018 project ``Analysis of STatic Analyses (ASTA)'', and \emph{Italian Ministry of University and Research}, under the PRIN2017 project no.\ 201784YSZ5 ``AnalysiS of PRogram Analyses (ASPRA)''.
	Francesco Ranzato has been partially funded by \emph{Facebook Research}, under a ``Probability and Programming Research Award''.
	%% The Appendices part is started with the command \appendix;
	%% appendix sections are then done as normal sections
	%% \appendix
	
	%% \section{}
	%% \label{}
	
	%% If you have bibdatabase file and want bibtex to generate the
	%% bibitems, please use
	%%
	\bibliographystyle{elsarticle-num} 
	\bibliography{references}
	
	%% else use the following coding to input the bibitems directly in the
	%% TeX file.
	
	%\begin{thebibliography}{00}
	
	%% \bibitem{label}
	%% Text of bibliographic item
	
	%\bibitem{}
	
	%\end{thebibliography}
\end{document}
\endinput
%%
%% End of file `elsarticle-template-num.tex'.

%%% Local Variables:
%%% mode: latex
%%% TeX-master: t
%%% End: